\documentclass[preprint2]{aastex}

\slugcomment{Version: \today}

\shorttitle{V1647 Orionis in Quiescence}
\shortauthors{Aspin, Beck, \& Reipurth}

\begin{document}

\title{V1647 Orionis: One Year into Quiescence}

\author{Colin~Aspin}

\affil{Institute for Astronomy, University of Hawaii,\\
  640 N. A'ohoku Place, Hilo, HI 96720 \\
  {\it caa@ifa.hawaii.edu}}

\author{Tracy~L.~Beck}
 \affil{Gemini Observatory, 670 N. A'ohoku Place, Hilo, HI 96720 \\
  {\it tbeck@gemini.edu}}

\author{Bo Reipurth}
\affil{Institute for Astronomy, University of Hawaii,\\
  640 N. A'ohoku Place, Hilo, HI 96720 \\
  {\it reipurth@ifa.hawaii.edu}}

\begin{abstract} 

We present new optical, near-IR, and mid-IR observations of the young   eruptive variable star V1647~Orionis that went into outburst in late   2004 for approximately two years.  Our observations were taken one year after the star had faded to its pre-outburst optical brightness and show that V1647~Ori is still actively accreting circumstellar material.  We compare and contrast these data with existing observations of the source from both pre-outburst and outburst phases.  From near-IR spectroscopy we identify photospheric absorption features for the first time that allow us to constrain the classification of the young star itself.  Our best fit spectral type is M0$\pm$2 sub-classes with a visual extinction of 19$\pm$2 magnitudes and a K-band veiling of r$_K\sim$1.5$\pm$0.2.  We estimate that V1647~Ori has a quiescent bolometric luminosity of $\sim$9.5~L$_{\odot}$ and a mass accretion rate of $\sim$1$\times$10$^{-6}$~M$_\odot$~yr$^{-1}$.  Our derived mass and age, from comparison with evolutionary models, are  0.8$\pm$0.2~M$_{\odot}$ and $\lesssim$~0.5~Myrs, respectively.  The presence towards the star of shock excited optical [S~II] and [Fe~II] emission as well as near-IR H$_2$ and [Fe~II] emission perhaps suggests that a new Herbig-Haro flow is becoming visible close to the star.

\end{abstract}


\section{INTRODUCTION}

A significant event in the recent history of star formation studies occurred in late 2003 when the young pre-main sequence star V1647~Orionis went into outburst.  This eruption, and the associated one hundred fold increase in optical brightness, resulted in the appearance of a monopolar reflection nebula now known as ''McNeil's Nebula'' after the amateur astronomer, Jay McNeil, who discovered it (McNeil 2004).  The star and nebula remained bright for approximately 18 months before fading rapidly over a six month period.   V1647~Ori was found to be again close to its pre-outburst (Sloan Digital Sky Survey, SDSS) optical brightness in early 2006.  A widely accepted interpretation of such eruptions is one involving a heated disk and a `mass dumping' episode from a circumstellar disk onto a stellar photosphere.  The five magnitude increase in optical brightness seen in V1647~Ori has been attributed to both the addition of significant accretion luminosity and a partial dust clearing due to the high-velocity wind (up to 600~km~s$^{-1}$) emanating from the star/disk engine (Reipurth \& Aspin 2004; McGehee et al. 2004).   

Both star and nebula have been widely studied from the period soon after the outburst occurred through to its fading to its pre-outburst brightness.  In our discussions below, we refer to the period before November 2004 as the {\it pre-outburst phase} (Brice\~no et al. 2004), the period from November 2004 to February 2006 as the {\it outburst phase}, and the period from February 2006 onwards as the {\it quiescent phase} of V1647~Ori.  Observations spanning all spectral regimes, from x-ray into the optical to the near-IR and on to the far-IR/sub-mm/radio, have been published by many authors in a large number of papers.  Readers are referred to these for detailed background information on V1647~Ori and McNeil's Nebula.  These papers are, in alphabetical order: \'Abr\'aham et al. (2004 - the pre-outburst infrared characteristics), Acosta-Pulido et al. (2007 - optical and near-IR imaging photometry, near-IR spectroscopy and polarimetry during the outburst, henceforth AP07), Andrews, Rothberg, \& Simon (2004 -- mid-IR and sub-mm observations during the outburst), Aspin et al. (2006 -- an historical study of previous outbursts), Brice\~no et al. (2004 -- the optical outburst history, pre-outburst and during the outburst), Fedele et al. (2007a,b -- optical and near-IR properties during the outburst), Gibb et al. (2006 -- near-IR spectroscopy during the outburst), Grosso et al. (2005 -- x-ray observations during the outburst), Kastner et al. (2004 -- x-ray emission pre-outburst and during the outburst), Kastner et al. (2006 -- x-ray evolution during the outburst), K\'osp\'al et al. (2005 -- optical photometry during the outburst), McGehee et al. (2004 -- optical and near-IR photometry pre-outburst and during the outburst), Mosoni et al. (2005 -- mid-IR interferometric observations during the outburst), Muzerolle et al. (2005 -- {\it Spitzer} observations during the outburst), Ojha et al. (2004 -- near-IR imaging during the outburst), Ojha et al. (2006 -- optical photometry and spectroscopy, and near-IR imaging during the outburst), Reipurth \& Aspin (2004 -- optical and near-IR imaging photometry and spectroscopy during the outburst), Rettig et al. (2005 -- high spectral resolution near-IR observations during the outburst), Semkov (2004 -- optical photometry during the outburst), Semkov (2006 -- optical light curve during the outburst), Tsukagoshi et al. (2005 -- mm continuum observations during the outburst), Vacca, Cushing, \& Simon (2004 -- near-IR spectroscopy during the outburst), Vig et al. (2006 -- radio observations during the outburst), and Walter et al. (2004 -- optical photometry and spectroscopy during the outburst). 

The significance of eruptive outbursts, similar to the one undergone by V1647~Ori lies in the fact that these periods are considered times when mass accretion rates increase dramatically (Hartmann \& Kenyon 1996, henceforth HK96).  During periods of outburst, mass accretion rates have been directly observed to increase by up to several orders of magnitude from a pseudo-steady-state value of $\sim$10$^{-9~to~-7}$~M$_\odot$~yr$^{-1}$ for classical T~Tauri stars (Hartmann et al. 1998) to $\sim$10$^{-6~to~-5}$~M$_\odot$~yr$^{-1}$ for EXor variables, and as high as $\sim$10$^{-4}$~M$_\odot$~yr$^{-1}$ for the extremely energetic FUor variables (Hartmann \& Kenyon 1985).  FUors and EXors are so named after the prototypes of their classes, namely FU~Orionis and EX~Lupi, respectively (Herbig 1989).  

Whether all young stars undergo EXor and/or FUor eruptions and, if so, whether these two types of outbursts events are fundamentally the same (on perhaps different scales) or different (in terms of the mechanism involved and the triggering of the eruption) is still a matter of debate.  EXor eruptions are distinct from FUor eruptions in that they are much shorter lived, months to years rather than decades to centuries, and have been empirically determined to be repetitive (Herbig 1989; 2007).  For example, EX~Lupi itself has gone into outburst on several occasions, with the last being in 1993--94 (Lehmann et al. 1995, Herbig et al. 2001).  V1647~Ori is likely another example of an EXor since its recent activity lasted only two years and it was previously observed in outburst from 1966--67 (Aspin et al. 2006) on archival photographic plates.  Mechanisms that have been proposed to initiate an eruptive event, be it EXor-like or FUor-like, include: thermal disk instabilities resulting in a runaway accretion condition (Bell \& Lin 1994), periodic overloading of the inner regions of the circumstellar disk and subsequent magnetic collapse (HK96), and the close approach of a companion in an eccentric orbit disturbing the stability of the inner regions of the disk (Bonnell \& Bastien 1992). 

In this paper, we present new optical, near-IR, and mid-IR observations of V1647~Ori and McNeil's Nebula taken approximately one year after the source reached its pre-outburst optical brightness.  Below, we compare and contrast our observations with those taken during the pre-outburst and outburst phases of the eruption and attempt to better understand the nature and characteristics of the underlying young star and its circumstellar environment.

\section{OBSERVATIONS \& DATA REDUCTION}

In order to study the on-set of the quiescent phase of V1647~Ori, we undertook an observing campaign involving optical, near-IR, and mid-IR instrumentation.  Table~1 relates the log of observations obtained together with associated photometric values derived from those datasets.  All the optical and  mid-IR observations were acquired on the ``Frederick C. Gillett'' Gemini North 8-meter telescope on Mauna Kea, Hawaii during the period UT February 13 to April 5 2007.  At this time, V1647~Ori had returned to its pre-outburst quiescent brightness for about one year.  Additionally, near-IR images and spectroscopy were acquired in February 2007 using the University of Hawaii 2.2m and the NASA IRTF 3.5m telescopes, respectively, both also located on Mauna Kea.

\subsection{Optical Imaging \& Spectroscopy}
\begin{figure*}[tb] 
\epsscale{1.0}
\plotone{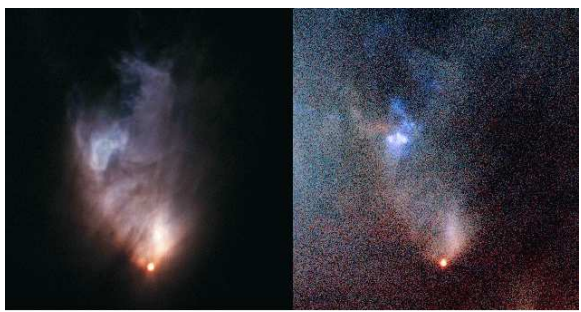} 
\caption{Two images of V1647~Ori and McNeil's Nebula.  The left image is a composite of g', r', and i' images taken on UT February 14, 2004.  The right image is a composite of r', i', and z' images was taken on UT February 22 2007.  North is at the top, East to the left in both images.  The images are approximately 70$''$ by 80$''$ in size at a pixel scale of 0.144$''$/pixel.  For comparison, V1647~Ori, at the apex of the nebula, is $\sim$5.6 magnitudes brighter in the Feb 2004 image taken a few months after the initial outburst.
\label{optcolim}}

\end{figure*}

Optical imaging data using Gemini North and GMOS-N (Davies et al. 1997) were obtained under Gemini queue program ID GN-2007A-Q-33 on UT February 22, 2007.  Images in the SDSS g', r', i', and z' filters were acquired in queue mode with an exposure time of 10 minutes per filter.  The data were reduced with the Gemini IRAF data reduction {\tt gmos} package (v1.9) using standard procedures for bias subtraction, flat-fielding (using master twilight flats), cosmic ray rejection, and image combination. Photometric calibration was performed using a photometric sequence of six field stars from SDSS images of the region.  Since this is a region of active star formation with the likelihood of some stars being intrinsically variable, we considered using as many as six field stars as calibrators to be an acceptable way to ensure insensitivity to such temporal changes.  A detailed discussion of the calibration sequence definition and properties will be presented in a subsequent paper on our two year outburst phase Gemini/IRTF optical, near-IR, and mid-IR monitoring program on V1647~Ori (Aspin \& Reipurth, in prep.).  The GMOS CCDs were binned 2$\times$2 to match the natural seeing of $\sim$0$\farcs$55.  A 'true-color' image of the region containing V1647~Ori, made from these data, is shown in Fig.~\ref{optcolim}. Also shown is a similar image taken some three years earlier during the outburst phase (February 2004).

Additionally, optical spectroscopy of V1647~Ori was obtained on UT February 21, 2007, again using Gemini North and the GMOS-N in queue mode.  We utilized the R400 grating with $\lambda _c$=7000~\AA\ and a 0$\farcs$75 wide long-slit.  This resulted in a spectral resolution of R$\sim$1280 and wavelength coverage of $\sim$5000--9150~\AA.  Seven 1800s exposures were taken giving a total on-source exposure time of 3.5 hours.  V1647~Ori was moved along the slit between each observation and, at the time of our optical spectroscopy, it had an r' magnitude of 23.26$\pm$0.15.  The data were again reduced using the Gemini IRAF {\tt gmos} package (v1.9) using flat-field and arc images from the facility calibration unit, GCAL.  In order to obtain the best sky-line subtraction in the far-red, we subtracted pairs of reduced 1800s images before residual sky-line subtraction and optimal point-source extraction.  These steps were performed using the Starlink FIGARO package. The seven extracted spectra of V1647~Ori were inter-compared to eliminate spurious noise spikes before combination.  The combined spectrum was flux calibrated using observations of the spectrophotometric standard star Feige~34.  Fig.~\ref{optspecplot} shows the final optical spectrum of V1647~Ori with all significant atomic lines identified.

\subsection{Near-IR Imaging \& Spectroscopy}
We obtained near-IR images of V1647~Ori on UT February 13, 2007 on the University of Hawaii 2.2~m telescope on Mauna Kea using ULBCAM (Hall et al. 2004).  This instrument consists of four Hawaii-2 1--5~$\mu$m arrays but operates in the J and H band only due to thermal background issues.  The field of view of the camera is 17$'\times$17$'$ with a pixel scale of 0$\farcs$25. A standard ``Mauna Kea'' H-band filter was used for the observations and the total on-source exposure time was 6 minutes.  Target images were dithered and median filtered to obtain sky images and flat-field images were taken by observing the inside of the telescope dome with lights on and off. Data reduction was performed using the Gemini IRAF {\tt quirc} data reduction package (v1.9).  For photometric calibration we used 2MASS sources within the field observed.  Fig.~\ref{himage} shows the region immediately surrounding V1647~Ori from these H-band images.  At the time of the observations, V1647~Ori had an H-band magnitude of 11.96$\pm$0.1 and is labeled in Fig.~\ref{himage} together with LkH$\alpha$~301 and HH~22-IRS (Muzerolle et al. 2005).

We used the NASA IRTF and the near-IR spectrometer SpeX (Rayner et al. 2003) to obtain near-IR 1--4.1~$\mu$m spectroscopy of V1647~Ori on UT February 22, 2007.  We observed using the SXD and LXD grisms with total exposure times of 1600 and 1200 seconds, respectively.  The data were reduced using the SpeXtools software package (Cushing, Vacca, \& Rayner 2004). An observation of an A0~V star close to the airmass of the target was used as a telluric calibrator and applied using the template fitting method defined in Vacca, Cushing, \& Rayner (2003).  V1647~Ori was nodded along the 0$\farcs$8 slit in an ABBA pattern to allow good sky definition and subtraction.  The final reduced and combined spectrum was flux calibrated using the H-band magnitude (H=11.96) from our ULBCAM imaging taken 9 days prior to eliminate slit-losses during the variable seeing conditions encountered.  The conversion of H magnitude to flux was performed using the Tokunaga \& Vacca (2005) values, for which a zero H magnitude star has a flux of F$_{\lambda}$=1.18~10$^{-9}$~W~m$^{-2}$~$\mu$m$^{-1}$. The combined SXD and LXD spectrum of V1647~Ori is presented in Fig.~\ref{nirspecplot}, where all major absorption and emission lines and bands are identified.  An expanded view of the 2.1--2.35~$\mu$m region of the spectrum of V1647~Ori is shown in Fig.~\ref{nirspecplot2}.

\subsection{Mid-IR Imaging \& Spectroscopy}
On UT March 18, 2007 we obtained Gemini North Michelle images of V1647~Ori in queue mode using the N' ($\lambda~_c$=11.2~$\mu$m) and Qa ($\lambda~_c$=18.5~$\mu$m) filters.  The total on-source exposure time was 100 seconds in each filter.  The chopped and nodded images were processed using the Gemini IRAF {\tt midir} package (v1.9), specifically task {\tt msreduce}.  Photometric calibration was performed using observations of the bright photometric standards 56~Ori and HD37160.  Since V1647~Ori appeared in both images as a point-source with N' and Qa band full width half maximum seeing of 0$\farcs$48 and 0$\farcs$66, respectively, they are not explicitly shown here.  From these images, however, we find that V1647~Ori had flux values of 0.23~Jy at 11.2~$\mu$m and 0.44~Jy at 18.5~$\mu$m. 

Additionally, mid-IR spectroscopy covering the 8--14~$\mu$m wavelength regime (the N-band) was obtained on Gemini North using Michelle with the 'low~N' grating on UT April 5, 2007.  The slit used was 0$\farcs$4 wide and resulted in a spectral resolution of R$\sim$200.  The on-source exposure time was 400 seconds and chopping along the long-slit ($\pm$10$''$) was employed.  A spectroscopic standard star (BS2473, spectral type G8~Ib) was observed immediately after the science observation and used to remove telluric features from the target spectrum.  Spectroscopic reduction was performed using the Gemini IRAF {\tt midir} package (v1.9) to give a reduced 2D spectral image, and the Starlink FIGARO package was used for optimal spectroscopic extraction and telluric correction. 

\section{RESULTS}

\subsection{The Optical Appearance}
In Fig.~\ref{optcolim} we compare V1647~Ori and McNeil's Nebula on images taken in February 2004 and three years later in February 2007. The images are derived from three optical passbands, the left image (February 2004) uses the g', r', and i' images while the right image (February 2007) uses the r', i', and z' data.  Note that the optical r'-band flux from V1647~Ori is $\sim$170$\times$ smaller in February 2007 than in February 2004, the latter being 2--3 months after the initial outburst.  We also note that in February 2007, some 12 months after V1647~Ori had returned to its pre-outburst brightness, we barely detect the reflection nebulosity and now see the HH~22 region (blue, to the north and slightly east of V1647~Ori) much more distinctly.  Even though the reflection nebula is very faint, all the physical features of it are still detected in the February 2007 image, specifically the fan-shaped region extending from the star to the north-west, the dark bands crossing the nebula on the eastern side of the monopolar lobe, the right-angled feature directly north of V1647~Ori, and the small nebulous blob a few arcseconds north-east.  Since we are observing the nebula in reflection (for example, see the polarization map of AP07), the brightness of the nebula should approximately mirror the brightness of V1647~Ori itself with appropriate delays for light-travel time (AP07).  We conclude, therefore, that the under-lying structure observed in the reflection nebula is quasi-static and independent of the eruption and brightness changes that have taken place.

Qualitatively, the scenario for the rapid brightening and fading of McNeil's Nebula is as follows.  The initial explosive eruption blew away and/or sublimated considerable obscuring dust along both our direct line-of-sight and the line-of-sight from V1647~Ori into the nebula (Reipurth \& Aspin 2004; Walter et al. 2004).  This was the direct result of both the initiation of an intense stellar wind and the enhancement to the UV/optical flux from the star due to the increased accretion rate.  We note that the bolometric luminosity, L$_{bol}$, of V1647~Ori increased from L$_{bol}\sim$3.5~L$_{\odot}$ pre-outburst to L$_{bol}\sim$50~L$_{\odot}$ during outburst (Andrews, Rothberg, \& Simon 2004) due to the addition of enhanced accretion luminosity (L$_{acc}$).  The addition of the L$_{acc}$ component and the blow out/sublimation of surrounding dust resulted in the brightening of V1647~Ori and hence, via reflection, McNeil's Nebula.   The reverse is also likely true, i.e. the decline of the L$_{acc}$ component and the slowing of the intense stellar wind resulted in the condensation of dust restricting the direct flux and the flux propagating into the nebula.  Dust sublimation was recently discussed by AP07 who commented on its reversibility i.e. sublimation during the outburst, and re-condensation after fading, and how such behavior was consistent with the documented evidence of at least one previous outburst (Aspin et al. 2006). 

It has been proposed that the existence and shape of McNeil's Nebula is the result of either a remnant wide-angle molecular outflow creating a cavity in the ambient molecular cloud material near V1647~Ori (Reipurth \& Aspin 2004), or that such nebulae are the result of space motion of the young star through the ambient molecular cloud (Goodrich 1987).  In the former case, an opposing outflow lobe would extend into the molecular cloud (since the visible lobe is tilted towards us -- AP07) and hence be hidden from view by cloud material.  As yet, there is no observational evidence for an active or remnant molecular outflow emanating from V1647~Ori (Lis, Menten, \& Zylka 1999, Andrews, Rothberg, \& Simon 2004).  Similarly, there has been no observational determination of space motion associated with V1647~Ori. 

\subsection{The Optical Spectrum}
\begin{figure*}[tb] 
\epsscale{0.9}
\plotone{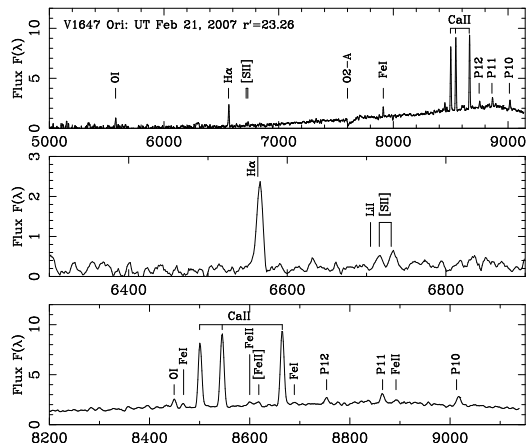} 
\caption{Optical spectroscopy of V1647~Ori taken on Gemini North using 
GMOS UT 22 February 2007.  The R400 grating was used with a $\lambda _c$=7000~\AA.  This resulted in a spectral resolution, R, of $\sim$1280.  The top panel shows the complete optical spectrum of V1647~Ori.  Note, the turnover in flux at 8900~\AA\ is an artifact of the data reduction.  The middle panel is a zoom of the region around H$\alpha$, while the bottom  panel is a zoom around the Ca~II triplet lines.  Flux, F$_{\lambda}$, is in W~m$^{-2}$~~$\mu$m$^{-1}$ and has been normalized to unity at 8000~\AA\ by division by 5.1$\times$10$^{-19}$.
\label{optspecplot}}

\end{figure*}

\begin{figure*}[tb] 
\includegraphics*[angle=-90,scale=0.75]{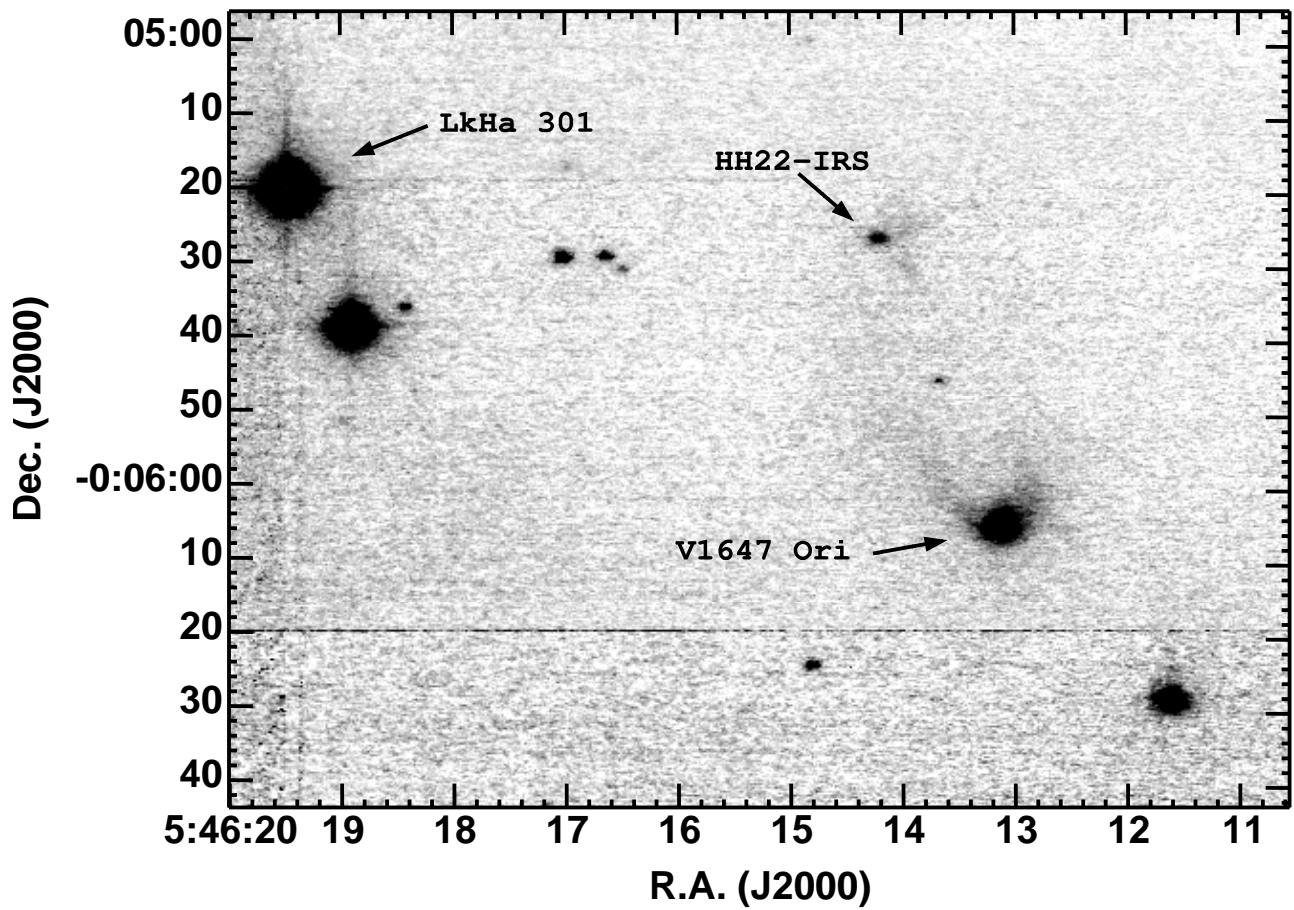} 
\caption{A near-IR H band image of the region containing V1647~Ori and McNeil's
Nebula.  This image was taken on the University of Hawaii 2.2m telescope on 
UT 13 February 2007.  V1647~Ori, the HH~22-IRS source, and Lk~H$\alpha$~301 are indicated. V1647~Ori has an H-band magnitude of 11.96. 
\label{himage}}

\end{figure*}

In Fig.~\ref{optspecplot} we present the optical spectrum of V1647~Ori taken in February 2007.  The top of the three panels shows the complete optical spectrum from 5000~\AA\ to 9150~\AA.  We note that at the time this spectrum was taken, V1647~Ori had an optical brightness of r'=23.3$\pm$0.1.  The spectrum seen is that of an object with a red continuum super-imposed with a variety of emission features.  The brightest emission lines are those from the Ca~II triplet at 8498, 8542, and 8662~\AA.  A pronounced H$\alpha$ emission feature, together with weak high-order H~I Paschen series P10, P11 and P12 lines, are also present.  Weaker emission lines are present also, specifically [S~II] at 6716 and 6731~\AA, O~I at 5579 and 8449~\AA, [Fe~II] at 8617~\AA, very weak Fe~I lines at 7912, 8467, and 8689~\AA, and weak Fe~II at 8600~\AA.  The absorption feature at 7600~\AA\ is the telluric O$_2$-A band. The middle and bottom panels of Fig.~\ref{optspecplot} show expanded views of the optical spectrum near H$\alpha$ (middle) and the Ca~II triplet (bottom).  The Li~I absorption line at 6707~\AA\ is indicated, however, it is not clear that it is really detected.  Table~2 shows the equivalent widths and line fluxes for the optical (and near-IR) emission features.

\subsubsection{Shock-excited [S~II] emission}
An interesting aspect of the optical spectrum is the presence of weak [S~II] emission lines with equivalent widths of --16.3~\AA\ (6716~\AA) and --22.0~\AA\ (6731~\AA).  Such [S~II] emission features were also detected by Fedele et al. (2007a) in their January 2006 observations. They interpreted these as possible evidence for the presence of a Herbig-Haro (HH) object close to the star.  We shall consider this interpretation further below.

We have made a careful study of the GMOS two-dimensional spectroscopic images to examine whether the [S~II] emission is present directly on the location of V1647~Ori or is also seen at other spatial positions along the long-slit used.  Our findings indicate that there is no [S~II] emission at all at any other location along the 90$''$ long-slit and we conclude that the shock-excited [S~II] emission lines are directly associated with the circumstellar region of V1647~Ori and are not due to foreground or background shocked nebulosity.  

What evidence is there to support the interpretation that the [S~II] lines originate in a shock-excited HH object?  First, Eisl\"offel \& Mundt (1997) considered that HH~23, located 155$''$ north of V1647~Ori and close to the axis of McNeil's Nebula, originated from V1647~Ori.  Second, as well as [S~II], the optical and near-IR spectrum of V1647~Ori shows weak [Fe~II] emission at 8617~\AA\ and 1.644~$\mu$m, and weak H$_2$ v=1-0~S(1) emission at 2.122~$\mu$m. Some of these lines are often associated with shock-excited HH objects (Gredel 1994, 1996; Hamann et al. 1994).  Third, it is becoming more widely recognized that there very likely exists a relationship between the creation of an HH object and the outburst phenomenon (Dopita 1978; Reipurth 1989).  

If we assume that HH~23 was ejected from V1647~Ori in perhaps an earlier outburst, then, with an inclination of the ejection axis with respect to the line-of-sight of $\sim$60 degrees (AP07) and assuming a typical HH object space velocity of ~$\sim$200~km~s$^{-1}$, HH~23 would have been ejected from V1647~Ori approximately 2000 years ago for a source distance of 450~pc.  However, it is, as yet, not definitively proven that HH~23 originates from V1647~Ori.  In Fig.~\ref{hh23} we show the region containing V1647~Ori, HH~22, and HH~23 imaged in narrowband [S~II] using GMOS-S on the Gemini South telescope on UT October 10 and 11, 2005.   This deep image clearly shows shock-excited nebulosity between V1647~Ori/McNeil's Nebula and HH~23 which resembles an intermediate bow shock with a trailing extended tail.  These features were identified as HH23~C by Eisl\"offel \& Mundt (1997).  Close inspection of Fig.~\ref{hh23}, suggests that it is equally conceivable that HH23~C and, in fact, the main HH~23 knots (A and B) could have originated from HH~22.  Within the HH~22~A shock-excited nebulosity lies a deeply embedded young Class~I protostellar source first detected by {\it Spitzer} (Muzerolle et al. 2005).  We henceforth refer to this source as HH22-IRS and identify its location in Fig.~\ref{hh23} with a white cross.  Muzerolle et al. (2005) considered that HH22-IRS drives the east--west HH~22 collimated flow and additionally noted that this flow is approximately orthogonal to the HH~23 flow.  If HH22-IRS drives both the HH~22 and HH~23 flows then it is perhaps a young binary system in which each component drives a (non-aligned) collimated flow.  Proper motion and radial velocity studies of both flows may clarify their origin.

If the presence of shock-excited optical and near-IR emission lines does indicate the presence of a new HH object close to V1647~Ori, when was this HH flow created?   We have investigated the possibility that these shock-excited [S~II] lines were present before January 2006, and hence could have been created in the explosive outburst itself.  In this case, the [S~II] lines would have been present in spectra of V1647~Ori taken prior to January 2006 but were perhaps overwhelmed by the then 100$\times$ brighter continuum flux.  We note that in January 2006, V1647~Ori had faded significantly from it outburst maximum brightness of r'=17.4 (Reipurth \& Aspin 2004).  We have simulated observing the same line strength [S~II] emission features on such an enhanced continuum.  In addition to considerably reducing the visibility of the lines on the high continuum level, a factor 100$\times$ increase in continuum flux implies the addition of 10$\times$ the photon shot noise to the data.  Under these conditions the [S~II] equivalent widths become 100$\times$ smaller and the noise becomes 10$\times$ larger.  We measure the root-mean square (rms) noise in a continuum subtracted region of the V1647~Ori spectrum close to the [S~II] lines at 3.5$\times$10$^{-21}$~W~m$^{-2}$~$\mu$m$^{-1}$ while the peak signal in the [S~II] lines is 5$\times$10$^{-20}$~W~m$^{-2}$~$\mu$m$^{-1}$.  This implies a signal-to-noise ratio (S:N) of $\sim$14.  If we increase the photon shot noise by a factor $\times$10 the S:N drops to $\sim$1.4 and would mean that the [S~II] lines of the same strength would likely not be detected during the outburst phase.

What we can conclude therefore, is that [S~II] emission was present in our February 2007 spectrum of V1647~Ori as well as the January 2006 spectrum of Fedele et al. (2007a).  We have shown that the emission is  isolated to the point-source itself but cannot determine whether these emission lines have recently formed or if they were present in earlier spectra and swamped by the enhanced continuum from V1647~Ori.  Again, if we assume an HH space velocity of $\sim$200~km~s$^{-1}$ and an inclination of $\sim$60 degrees, a new HH object would take $\sim$13 years to move $\sim$1$''$ (for a distance of 450~pc).

The fact that [S~II] emission lines are present, however, implies that shocks exist in the immediate vicinity of the accreting young star.  Since the [S~II] 6716~\AA/6731~\AA\ line ratio is only weakly dependent on temperature, it can be used as a diagnostic of the electron density in the [S~II] emitting region.  Using the IRAF {\tt nebular.temden} routine and our February 2007 observed line ratio of $\sim$0.8, we estimate that the electron density, {\it n$_e$}, was $\sim$1400~cm$^{-3}$ for an assumed electron temperature, T$_e$, of 10$^4$~K (a value consistent with mild shock excitation).  If we vary T$_e$ over the range 5000--20000~K (the temperature range over which [S~II] emission can occur), a ratio of 0.8$\pm$0.1 gives a range of {\it n$_e$} of $\sim$1000--1800~cm$^{-3}$. This derived value is a factor $\sim$14$\times$ lower than the [S~II] critical density of $\sim$2$\times$10$^{4}$~cm$^{-3}$ (Bacciotti 2002).  Performing the same analysis for the Fedele et al. (2007a) data from January 2006, we find a [S~II] line ratio of $\sim$0.6$\pm$0.1 and an electron density value of {\it n$_e$}$\sim$4200~cm$^{-3}$ with a range for valid T$_e$ values of 3200--5300~cm$^{-3}$.  The Fedele et al. (2007a) value is perhaps a little larger than than found in February 2007 although the errors on the ratios are significant.  A difference in this line ratio would mean that conditions in the region where the shock-excitation occurs have evolved over the intervening year and could be interpreted as either physical electron density changes or changes in the local ionization fraction. There is some evidence that in HH flows, higher electron densities may occur closer to the exciting sources (Cohen \& Jones 1987; B\"uhrke, Mundt, \& Ray 1988; Heathcote \& Reipurth 1992; Beck et al. 2007).  If the above ratio decrease is real, then we can speculate that perhaps the shocked region has moved further away from the star. 

\subsubsection{H$\alpha$ emission}
The equivalent width of the H$\alpha$ emission in our spectrum is W$_{\lambda}$=--124$\pm$10~\AA, the large uncertainty being due to the very weak continuum flux present at the wavelength of H$\alpha$.  The observed W$_{\lambda}$ value is considerably larger than that last reported: Ojha et al. (2006) found W$_{\lambda}$=--40.75 on UT September 28, 2005.  This is also likely to be a result of the weak observed continuum flux.

The H$\alpha$ emission line flux in our spectrum is F$_{\lambda}$=1.0$\times$10$^{-18}$~W~m$^{-2}$ and we can compare this directly to the outburst phase H$\alpha$ line fluxes presented in AP07, Ojha et al. (2006), and Walter et al. (2004).  AP07 plot their own H$\alpha$ line fluxes, together with those of Ojha et al. (2006) and Walther et al. (2004), against time (their Fig.~12) from soon after the outburst was detected in early 2004 through late 2005.  Their plot shows a declining H$\alpha$ line flux during the outburst period, from a peak in early 2004 of $\sim$2.0$\times$10$^{-17}$~W~m$^{-2}$ to $\sim$4.0$\times$10$^{-18}$~W~m$^{-2}$ in late 2005.  Our quiescent phase value from early 2007 is lower still by a factor $\times$4 and follows the outburst trend.  As pointed out by AP07, H~I emission lines are often directly related to the accretion process.   This trend could represent declining accretion over the outburst to quiescent phase period although as we shall see below, changes in visual extinction could also affect the observed fluxes.  We defer a discussion of derived accretion rates during these phases to Section~3.7.3 where we consider the near-IR spectrum of V1647~Ori in relation to the models of Muzerolle et al. (1998, 2001). 

We additionally note that the H$\alpha$ line profile appears somewhat asymmetric in shape.  Specifically, the red wing appears truncated with respect to the blue wing and with respect to the profile of other emission lines (e.g. Ca~II).  Such a red-wing truncation has not, to our knowledge, been seen before in V1647~Ori.\footnotemark\footnotetext{A recent high spectral resolution   Keck~II/NIRSPEC near-IR spectrum of V1647~Ori taken by us shows a   similarly truncated Br$\gamma$ emission line.}  We can perhaps attribute this asymmetry to redshifted absorption from infalling cool neutral gas although higher spectral resolution optical data would be required before a conclusive decision can be made.

\subsubsection{Ca~II triplet emission}
In the quiescent-phase spectrum shown in Fig.~\ref{optspecplot}, the Ca~II triplet lines at 8498, 8542, and 8662~\AA\ have equivalent widths, W$_{\lambda}$, of --30.6$\pm$1~\AA, --35.4$\pm$1~\AA, and --34.2$\pm$1~\AA, respectively.  The ratio of these lines, specifically, $\sim$1.7:2:1.9, is somewhat different from that found by Walter et al. (2004) in spectra from the outburst period taken between February and April 2004.  They reported a Ca~II triplet ratio of $\sim$2:2:1, a pattern considered unique to T~Tauri and Herbig Ae/Be stars (Herbig \& Soderblom 1980; Hamann \& Persson 1990), and inconsistent with the expected ratio from (optically thin) transition strengths (gf) i.e. 1:9:5.  The 2:2:1 line ratio was also noted to be different from that produced by either simple optical depth effects or chromospheric models (Hamann \& Persson 1992).  We have compared the observations and model predictions presented in both Fig.~8 of Hamann \& Persson (1992) and in Fig.~3 of Herbig \& Soderblom (1980) with our derived ratios to investigate the physical conditions in which Ca~II lines arise.  We note that in the above papers, the quantities used are W$_{\lambda}$(8498)/W$_{\lambda}$(8542) and W$_{\lambda}$(8662)/W$_{\lambda}$(8542).  Our measured ratios of $\sim$1.7:2:1.9 therefore become 0.86$\pm$0.05 and 0.97$\pm$0.05, respectively.  A similar ratio was also found by Gibb et al. (2006) and Vacca et al. (2004) in their post-outburst spectra.  In Fig.~8 of Hamann \& Persson (1992), V1647~Ori would lie close to the loci of models for either i) a very high optical depth ($\tau\sim$1000) isothermal LTE slab, or ii) a somewhat lower optical depth isothermal non-LTE slab ($\tau\sim$50, but still optically thick).  The location of V1647~Ori is distinctly separated from both the majority of T~Tauri stars they observed (i.e. with mean ratios of $\sim$1,0.8), and chromospheric model values (i.e. 0.6,0.8).  Following Hamann \& Persson (1992), ratios close to unity as in our case, indicate that collisional decay is dominant over radiative decay and requires conditions with $\tau$n$_e\ge$10$^{13}$~cm$^{-3}$ where n$_e$ is the electron density.  It seems, therefore, that regardless of LTE or non-LTE conditions, in February 2007 the Ca~II triplet lines in V1647~Ori were extremely saturated and produced in a very high density medium.  This is also supported by the location of V1647~Ori in Fig.~3 of Herbig \& Soderblom (1980), where V1647~Ori would lie close to the optically thick LTE slab location.   At this location, several T~Tauri stars are found, specifically, GW~Ori (0.93,0.95), V380~Ori (0.88,0.87), V1121~Oph (0.85, 0.91), and RW~Aur (1.02,0.84).  We finally noting that it appears the Ca~II triplet ratio changed sometime during the post-outburst phase between March and November 2004 suggesting a fundamental physical change in the region where the Ca~II lines are emitted.

\subsection{The Near-IR Appearance}
Of primary interest in our H-band image shown in Fig.~\ref{himage} is the fact that V1647~Ori itself is nebulous and not just a point-source.  The nebulosity takes the form of curving tails extending from the star to the north-east and north-west.   These features were also observed in the K-band by Reipurth \& Aspin (2004), in J, H, and K by Ojha et al. (2004) and in J and K by AP07.  Our image shows somewhat more extensive nebulosity, particularly at very faint levels, reaching as far as HH~22-IRS. These curving 'cometary' tails most likely trace the cavity walls of a remnant outflow cavity (although, as we have noted, no such outflow has yet been found). However, it has also been suggested that such a cavity may be a remnant of infall to the circumstellar envelope (Whitney \& Hartmann 1993; Kenyon et al. 1993; and Hartmann, Calvet, \& Boss 1996).

Such cavities, be they from infall or outflow, have been modeled by Reipurth et al. (2000) and later Stark et al. (2006).  The purely geometric models of Reipurth et al. (2000) were applied to {\it HST} NICMOS images of Herbig-Haro energy sources and predicted the  cavity shape for different opening angles and inclinations.  A cursory comparison between the McNeil's Nebula optical image from February 2004 (Fig.~\ref{optcolim}) and their results suggests that an inclination with respect to the line-of-sight, $\theta$, of $\sim$60 degrees and cavity opening angle, $\alpha$, of $\sim$30 degrees is appropriate.  More quantitatively, applying formulae (1) for $\theta$ and (2) for $\alpha$ from Reipurth et al. (2000) we obtain values of $\theta\sim$65 degrees and $\alpha\sim$30 degrees.  

In the models of Stark et al. (2006), the cavity walls are assumed to be seen in reflected light.   They predict the near-IR outflow cavity geometry for a variety of physical parameters associated with both the circumstellar disk and an extended infalling envelope.  They also consider the effect of cavity inclination and the shape/form of the cavity e.g. a 'streamline' cavity with straight walls extending away from the central star, or a 'curved cavity' (as we see in V1647~Ori) on the derived morphology.  Although these models were created specifically to interpret {\it HST} NICMOS images of young stellar objects in Taurus (e.g. IRAS 04302+2247, DG~Tau~B, and CoKu~Tau/1), their use as a qualitative comparison to our V1647~Ori H-band image seems appropriate.  Of the NICMOS observations considered in Stark et al. (2006), those of CoKu~Tau/1 appear most similar in morphology to our near-IR data on V1647~Ori although the spatial scales considered are clearly much different.  The numerical model that best fits the CoKu~Tau/1 data involves a curving (bipolar) cavity with opening angle 20 degrees and inclination to the line-of-sight of 64 degrees, an infalling envelope with infall rate of 2$\times$10$^{-7}$~M$_{\odot}$yr$^{-1}$, and a disk with a 50~AU radius disk.  Clearly, we cannot make any claim that the best-fit parameters for CoKu~Tau/1 are appropriate for V1647~Ori, only that such parameters produce a model that qualitatively resembles the cavity walls observed. An important point, however, is that the models predict that the extent of the cavity wall illumination changes significantly with outflow axis inclination and that the most appropriate model for CoKu~Tau/1 has an inclination of 64 degrees, a value close to that derived by AP07.  

In addition to V1647~Ori, the H-band image shown in Fig.~\ref{himage} contains several other point-sources including the interesting near-IR object HH~22-IRS.  The nebulous extensions seen in the IRAC images of this source are also detected in our H-band images and likely take the form of cavity walls extending to the west from HH~22-IRS.  These cavity walls appear more streamlined than curved.  Higher spatial resolution, high signal:noise data may shed light on the origin of both types of cavities.  

\subsection{The Near-IR Spectroscopic Features}
\begin{figure*}[tb] 
\epsscale{1.0}
\plotone{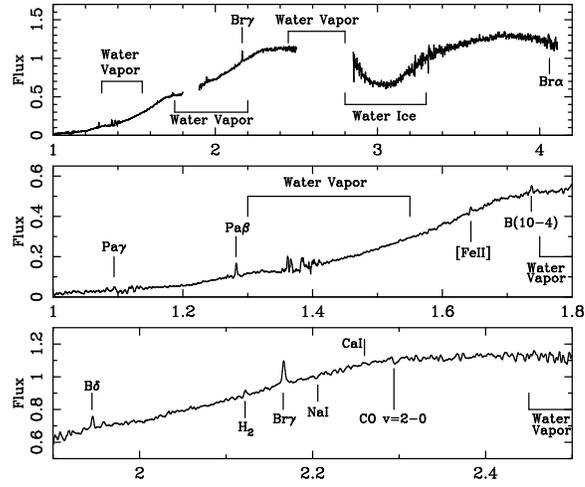} 
\caption{Near-IR spectroscopy of V1647~Ori taken on the NASA IRTF using 
SpeX on UT 22 February 2007.  Both the SXD and LXD grisms were used producing
X-dispersed spectra of the 1 to 2.5~$\mu$m and 2.5 to 4.1~$\mu$m region, respectively.
The top panel shows the complete spectrum while the middle and bottom panels
show the J$+$H and K band spectral regions, respectively.  The spectrum is
calibrated in units of F$_{\lambda}$ such that the flux at 1.65~$\mu$m is 
1.94~10$^{-14}$~W~m$^{-2}$~~$\mu$m$^{-1}$.  The spectrum was normalized to 
unity at 2.2~$\mu$m where the flux is 4.54~10$^{-14}$~W~m$^{-2}$~~$\mu$m$^{-1}$.
\label{nirspecplot}}

\end{figure*}

\begin{figure*}[tb] 
\epsscale{1.0}
\plotone{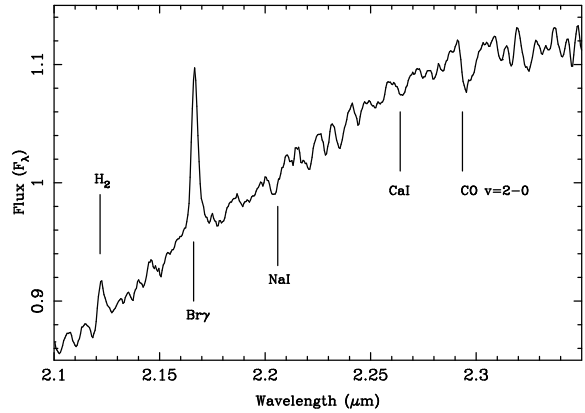} 
\caption{An expanded view of the 2.1--2.35~$\mu$m region of the near-IR 
spectroscopy of V1647~Ori. The spectrum is calibrated in units of 
F$_{\lambda}$ and normalized to unity at 2.2~$\mu$m where the flux is 
4.54~10$^{-14}$~W~m$^{-2}$~~$\mu$m$^{-1}$.  Note the presence of weak CO
bandhead absorption at 2.294~$\mu$m.
\label{nirspecplot2}}

\end{figure*}

\begin{figure*}[tb] 
\includegraphics*[angle=0,scale=0.75]{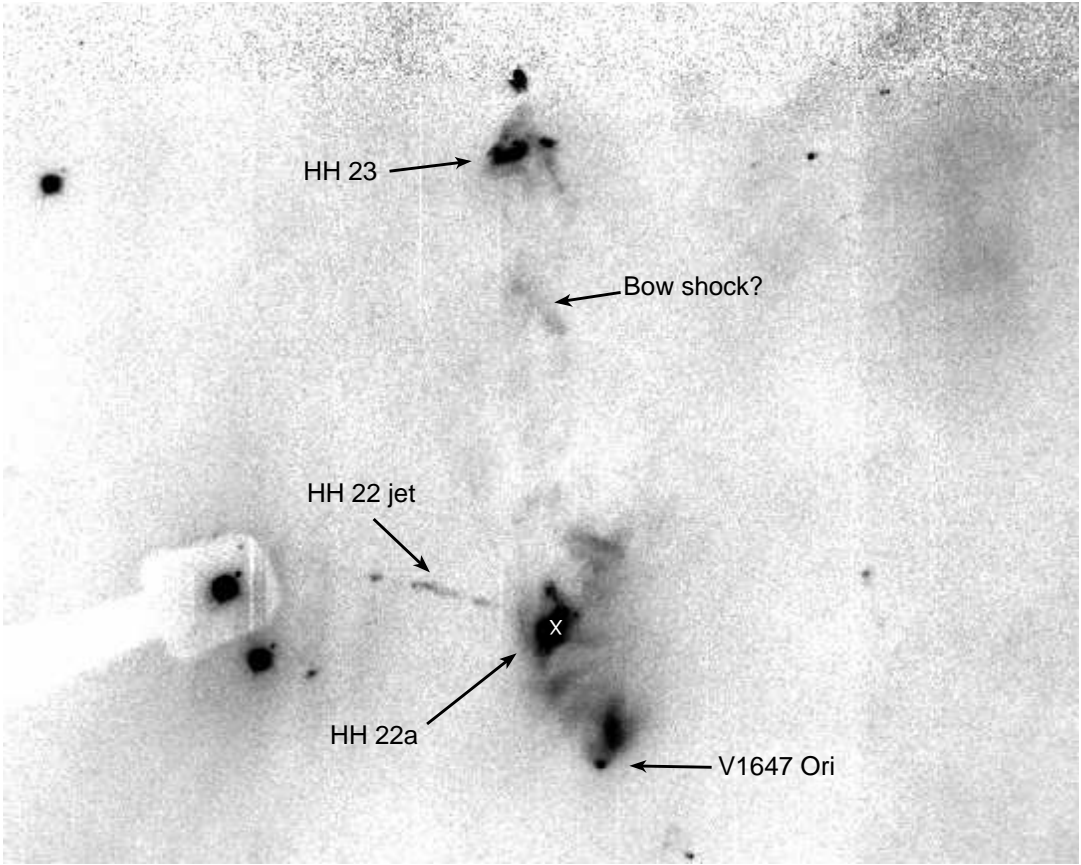} 
\caption{An optical narrowband [S~II] image of V1647~Ori, McNeil's Nebula, and the HH object HH~22 and HH~23.  North is at the top, east is to the left.  HH~23 is 155$''$ north of V1647~Ori and appears to originate from it.  A possible intermediate bow shock is marked as are the HH~22 jet, and the far-IR source detected by {\it Spitzer}, HH22-IRS (Muzerolle et al. 2004). 
\label{hh23}}

\end{figure*}

Fig.~\ref{nirspecplot} shows the near-IR spectrum of V1647~Ori from 1 to 4.1~$\mu$m taken with SpeX on the NASA IRTF telescope. The top panel shows the complete spectrum while the middle and bottom panels show expanded views of the J and H band (middle), and K band (bottom). Numerous spectral emission and absorption features are identified from both atomic and molecular transitions.  Inspection of Fig.~\ref{nirspecplot} shows that a year after the outburst had subsided, V1647~Ori shows broad water vapor absorption features (1.3--1.6~$\mu$m, 1.75--2.2~$\mu$m, and 2.35--2.5~$\mu$m), as well as deep water ice absorption (2.8--3.3~$\mu$m).  Particularly significant is the presence of weak CO overtone absorption (2.294~$\mu$m), weak atomic Na~I absorption (2.206~$\mu$m), and weak Ca~I absorption (2.264~$\mu$m).  These features are best seen in the expanded 2.1--2.4~$\mu$m spectrum in Fig.~\ref{nirspecplot2}.  In emission, we see relatively strong Br$\gamma$ (2.166~$\mu$m), weak H$_2$ v=1-0 S(1) (2.122~$\mu$m), weak [Fe~II] (1.275, 1.533, 1.644, and 1.809~$\mu$m), and several other H~I lines from the Paschen and Brackett series, the longest in wavelength being Br$\alpha$ (4.05~$\mu$m).

Of particular note is the evolution of the CO overtone features from 2.3--2.5~$\mu$m through the outburst.  In the near-IR spectra of Reipurth \& Aspin (2004) and  Vacca et al. (2004), taken in February and March 2004, respectively, the CO overtone bandheads were strongly in emission.  As the outburst progressed the CO emission bandheads faded and in the May 2006 spectrum of AP07 the 2.3--2.5~$\mu$m region was featureless and remained so in their August 2006 spectrum.  In our spectrum from February 2007, however, we are now seeing weak CO bandheads in absorption.   As we shall see below (Section 3.5), we believe that we are observing CO, Na~I and Ca~I absorption from the stellar photosphere of V1647~Ori.  

The Br$\gamma$ emission flux from V1647~Ori was also tracked by AP07.  They found an approximately linear decline in Br$\gamma$ flux from 2.2$\times$10$^{-16}$~W~m$^{-2}$ in March 2004 to 3.3$\times$10$^{-17}$~W~m$^{-2}$ in September 2006.  Our Br$\gamma$ flux of 2.3$\times$10$^{-17}$~W~m$^{-2}$ in February 2007 follows this trend.  A similar trend is found in both He~I (1.083~$\mu$m) and Pa$\beta$ (1.282~$\mu$m) emission with an approximately linear trend declining by factors $\sim\times$30 and $\sim\times$17, respectively, from March 2004 to February 2007.  

The broadest features present in the near-IR spectrum of V1647~Ori are due to water molecules.  Specifically, we see water vapor absorption features in the J, H, and K bands, and water ice absorption in the L band.  While the water ice absorption feature has been observed in V1647~Ori from the very first observations at 3~$\mu$m (Vacca et al. 2004), water vapor absorption is detected for the first time.  

We find that the water ice absorption depth and profile in our spectrum appears very similar to the profiles found in the spectra presented by Rettig et al. (2005) and Gibb et al. (2006). This perhaps suggests that the ice mantle grains were not significantly affected by the outburst itself.  

The presence of water vapor absorption suggests that we are observing a cool photosphere of spectral type later-M.  However, Shiba et al. (1993) suggested that a significant fraction of young low-mass stars show water vapor absorption in excess of their true spectral type.  This was interpreted as evidence for absorption in the extended atmosphere of a circumstellar disk.   Qualitatively the water vapor absorption we observe in V1647~Ori seems inconsistent with the atomic lines and molecular band absorption seen in the K band.  A more detailed discussion of this is given below where we apply stellar models to the observed spectrum to derive physical parameters associated with the young star itself and the circumstellar material.

\subsection{Near-IR spectral modeling}

The detection of neutral Na (2.206~$\mu$m) and Ca (2.265~$\mu$m) atomic absorption, and ro-vibrational molecular CO {\it v}=2-0 bandhead (2.294~$\mu$m) absorption  in the near-IR spectrum of V1647~Ori can be interpreted as evidence that we are detecting flux from a stellar photosphere.  However, it is also possible that these spectral features are created in a disk atmosphere.  The atmospheres of circumstellar disks are considered to be low-surface gravity environments similar in many ways to the atmospheres of cool supergiants.  In such atmospheres, molecular CO bandhead absorption has been found to be extremely strong with respect to neutral atomic Na and Ca absorption.  Examples of such disk-dominated near-IR spectra can be found in FUors where the 2$\mu$m spectrum is often relatively featureless besides deep, broadened CO absorption (see for example Fig.~1 in Reipurth \& Aspin 1997).  In V1647~Ori, unless the CO absorption is significantly diluted by CO emission, the strength of the Na and Ca lines with respect to the CO absorption suggests that the environment of the absorption region is more dwarf-like than supergiant-like.  However, we find it is not perfectly dwarf-like but  intermediate between dwarf and giant values.  This is consistent with V1647~Ori being a young, pre-main sequence dwarf star still contracting along Hayashi tracks.  In the discussion and analysis that follows, therefore, we assume that the Na, Ca, and CO absorption features are from the stellar photosphere of V1647~Ori but note that this assumption is not unambiguously determined.

By modeling the strength of photospheric absorption features, using, for example, spectral template stars, allows us to derive estimates of the underlying stellar type, the characteristics of the IR excess emission, and the extinction toward the photosphere.  We have first applied the modeling method of Prato, Greene \& Simon (2003) to the K-band spectra of V1647~Ori.  This is the spectral region where the aforementioned photospheric lines are found.  The spectrum was fit with a series of models of spectral type template stars which were veiled with infrared emission excess and reddened by extinction using the ISM extinction law:

\begin{equation}
A_{\lambda}~=~A_v[0.55/\lambda]^{1.6} 
\end{equation}

taken from Prato, Greene \& Simon (2003) where 0.55 and $\lambda$ are in microns.  The K-band veiling, r$_K$, is defined as the ratio of the infrared excess at 2.22~$\mu$m to the intrinsic stellar continuum flux.  The results of Muzerolle et al. (2003) and Beck (2007) show that IR excess can be described by blackbody emission with a single characteristic temperature and hence, for the purpose of the K-band models, we use an IR excess with a blackbody temperature, T$_{dust}$, of 1000~K.  A $\chi^2$ search for the optimum r$_K$ value and the best-fit A$_V$  was performed for a range of stellar template stars from K5 to M3.  The reader is referred to Prato, Greene \& Simon (2003) and Beck (2007) for more details of the models.  

\begin{figure*}[tb] 
\includegraphics*[angle=0,scale=0.8]{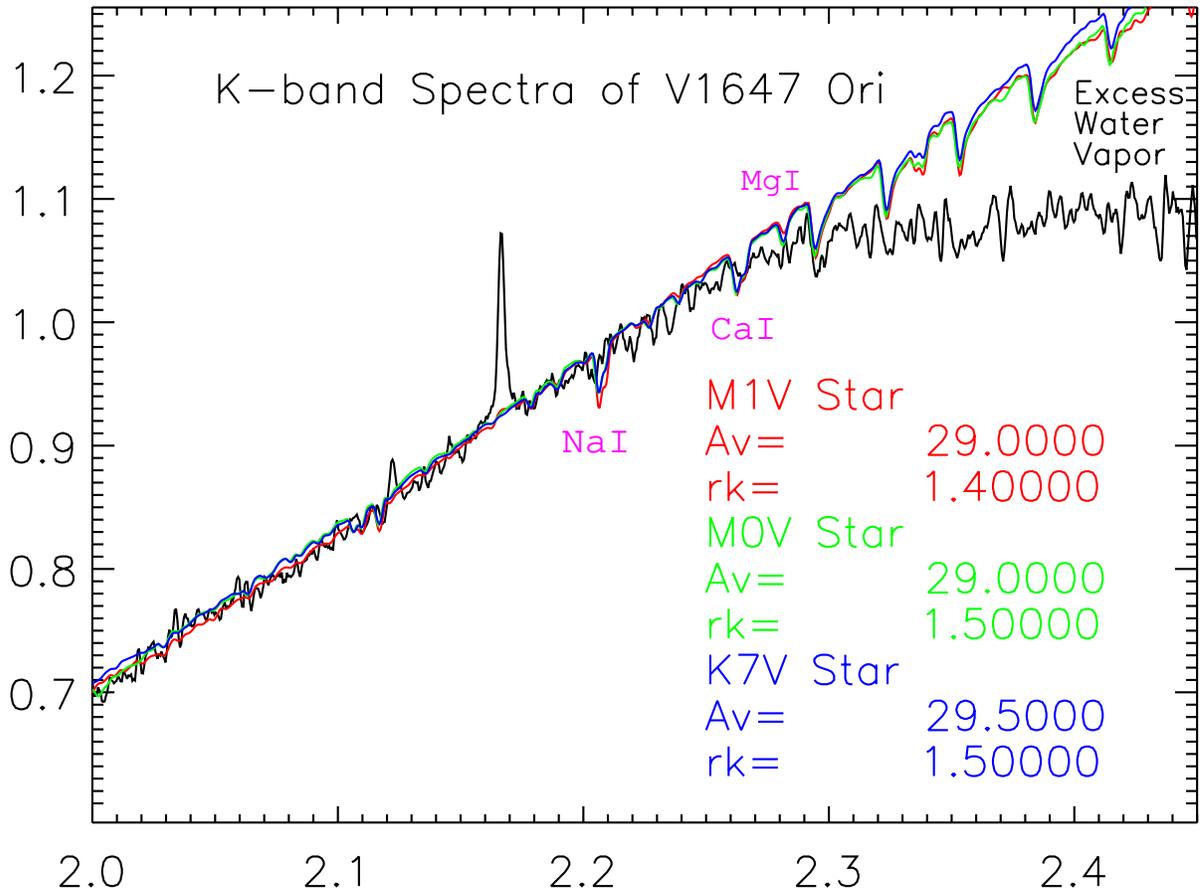} 
\caption{A plot of the K-band spectrum of V1647~Ori (black) overlaid by model fits using the method of Prato, Greene, \& Simon (2003).   Three different model fits are shown for spectral types M1~V (orange), M0~V (green), and K7~V (blue).  These produce the best overall fit to the data.  The derived A$_V$ and r$_K$ values for these models are 29 and 1.5, respectively.  The poor fit longward of 2.3~$\mu$m is due to the presence of excess water vapor absorption from the circumstellar environment. 
\label{TB_Figure1}}

\end{figure*}

In Fig.~\ref{TB_Figure1}, we plot the K-band spectrum of V1647~Ori (in black) with the best-fit models for three template stellar types,  K7~V, M0~V and M1~V.  Stellar types in the K5 to M2 range provide the best average fit to the detected photospheric features: stars with much earlier types (i.e. $\sim$K3 or earlier) have Mg~I (2.281~$\mu$m) absorption  which should have been detected at the same strength as Ca (2.265~$\mu$m), and stars with much later stellar types (i.e. $\sim$M3 and later) have much stronger Na~I (2.206~$\mu$m) absorption relative to Ca~I.  This is true for both dwarf stars and giants. The stellar templates were taken from the NASA IRTF SpeX stellar spectral library located on the IRTF web~site\footnotemark \footnotetext{http://irtfweb.ifa.hawaii.edu/~spex/spexlibrary/IRTFlibrary.html}.  From the K-band model fits, the best-fit values give A$_v\sim$29 magnitudes, and r$_k$=1.4--1.5.  However, early in the process of modeling the K-band spectra, it became apparent that the spectrum at wavelengths greater than $\sim$2.3~$\mu$m deviated significantly from the models which best-fit the photospheric absorption features (see Fig.~\ref{TB_Figure1}).  This deviation is best explained by excess absorption from water vapor often seen, for example, in FUors (Greene \& Lada 1996).  The presence of water vapor features significantly affects the extinction values determined by modeling the K-band alone.  Fitting the steep slope of the K-band spectra at $\lambda<$2.3~$\mu$m (made steeper by the 1.9--2.2~$\mu$m water vapor band) requires extinction value of $\sim$29 magnitudes.  This is a much larger A$_V$ than previously associated with V1647~Ori (i.e. 6--14 magnitudes, Aspin \& Reipurth 2004; Brice\~no et al. 2004; \'Abrah\'am et al. 2004; Andrews et al. 2004; Vacca et al. 2004, Ojha et al. 2006, AP07) and, since water vapor absorption occurs shortward of 2.2~$\mu$m also, we conclude that the K-band models are likely affected by the presence of water vapor and are therefore inadequate for the derivation of physical characteristics.

\begin{figure*}[tb] 
\includegraphics*[angle=90,scale=0.7]{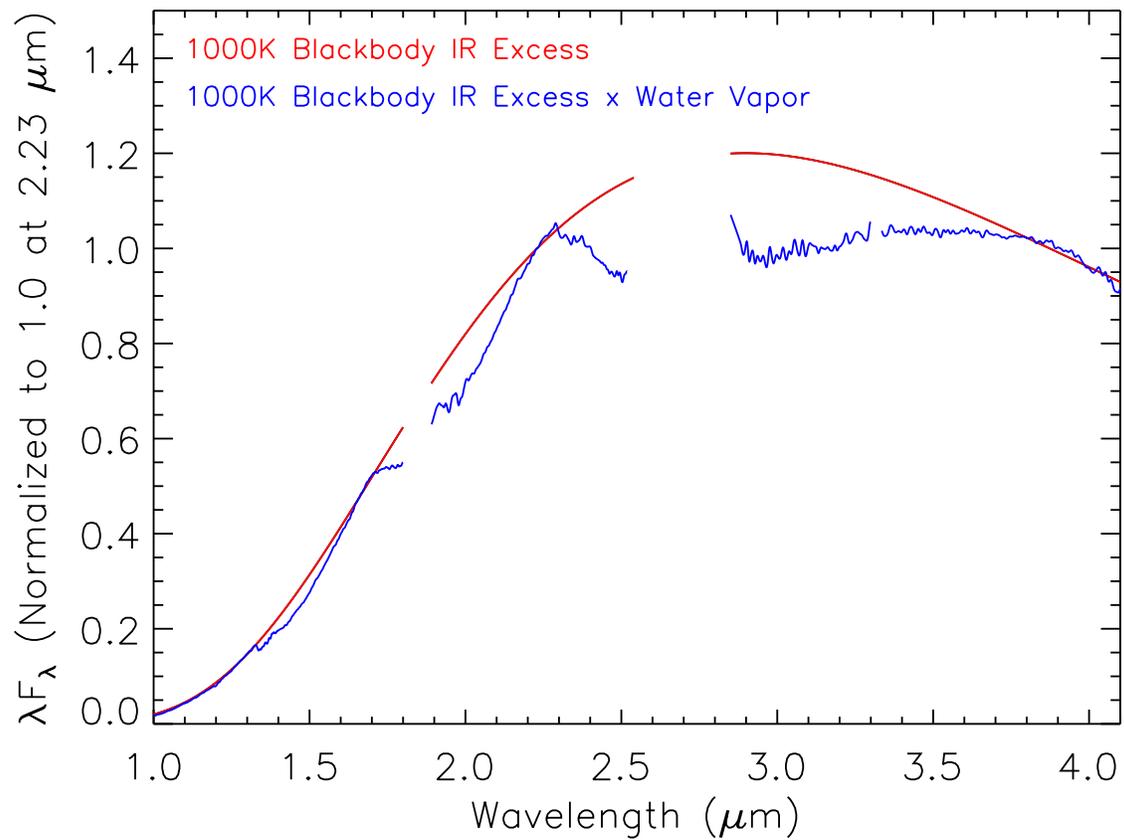} 
\caption{A plot of our adopted near-IR disk excess emission model.  The red curve shows the shape of the near-IR disk excess emission of temperature 1000~K.  In the blue curve, we have incorporated a rudimentary model for water vapor absorption by adding the shape of the water vapor features of an M7~V star.  This was achieved by {\it i)} removing all other spectral features from the M7~V data, {\it ii)} smoothing the data, and {\it iii)} removing the M7~V continuum shape by subtracting an appropriate black-body curve.
\label{TB_Figure2}}

\end{figure*}

To further investigate the above models validity, we have extended the near-IR spectral modeling of Prato, Greene, \& Simon (2003) and the version used by Beck (2007) to include the full 1.0--4.1~$\mu$m spectral range in the hope of placing more stringent constraints on both A$_V$ and r$_K$.  The extended synthesis model was adapted from the K-band spectral code to use an optical depth model which included water ice absorption, and an infrared excess model that included water vapor absorption.  The model of the observed 1--4.1~$\mu$m spectrum is defined as:

\begin{equation}
F_{obs} = {\it C} \times [(F_{int} + k(\lambda))e^{-\tau(\lambda)}]
\end{equation}

where F$_{int}$ is the intrinsic photospheric flux of the young star estimated from the spectral template star.  The wavelength dependent optical depth, $\tau(\lambda)$, has two components: first from the ISM extinction law as described in Eqn.~1 above, and second from an optical depth component, $\tau_{ice}$, derived directly from laboratory absorbance data for water ice (Gerakines et al. (1995; 1996\footnotemark{\footnotetext{Laboratory water ice absorbance data are available online at: http://www.strw.leidenuniv.nl/$\sim$lab}}).  The parameter {\it C} is a scaling constant to normalize the model to  unity at 2.22~$\mu$m.  The value k($\lambda$) is the wavelength dependent emission excess, which is described using black-body radiation with a temperature, T$_{dust}$, constrained by the required r$_K$ value at 2.22~$\mu$m determined by the K-band models (Muzerolle et al. 2003).  We have made an important modification to the Muzerolle et al. infrared excess model by incorporated an estimate of water vapor absorption.  To accomplish this, we have multiplied the blackbody curve that describes the infrared excess emission by a smoothed template model of water vapor absorption.  The results are presented in Fig.~\ref{TB_Figure2}.  The red curve is the blackbody IR excess with a temperature of 1000~K, and the blue curve is this blackbody multiplied by the water vapor absorption model.  The water vapor absorption model was constructed by {\it i)} removing all photospheric absorption features from a M7~V star (using the SpexTool ``xcleanspec'' routine -- Cushing, Vacca, \& Rayner 2004), {\it ii)} smoothing the cleaned spectrum using a 10-pixel boxcar, {\it iii)} dividing the smoothed spectrum by a black-body of temperature 2600~K to remove the M7~V spectral continuum shape, and {\it iv)} normalizing the flux of the water vapor model to unity at 2.22~$\mu$m.  An M7~V stellar spectrum was adopted since it best fit the water absorption observed in V1647~Ori.  Since water vapor characteristics are not derived by the fitting procedure and only allow us to obtain more meaningful values of the free-parameters, we deemed this methodology acceptable. 

\begin{figure*}[tb] 
\includegraphics*[angle=0,scale=1.0]{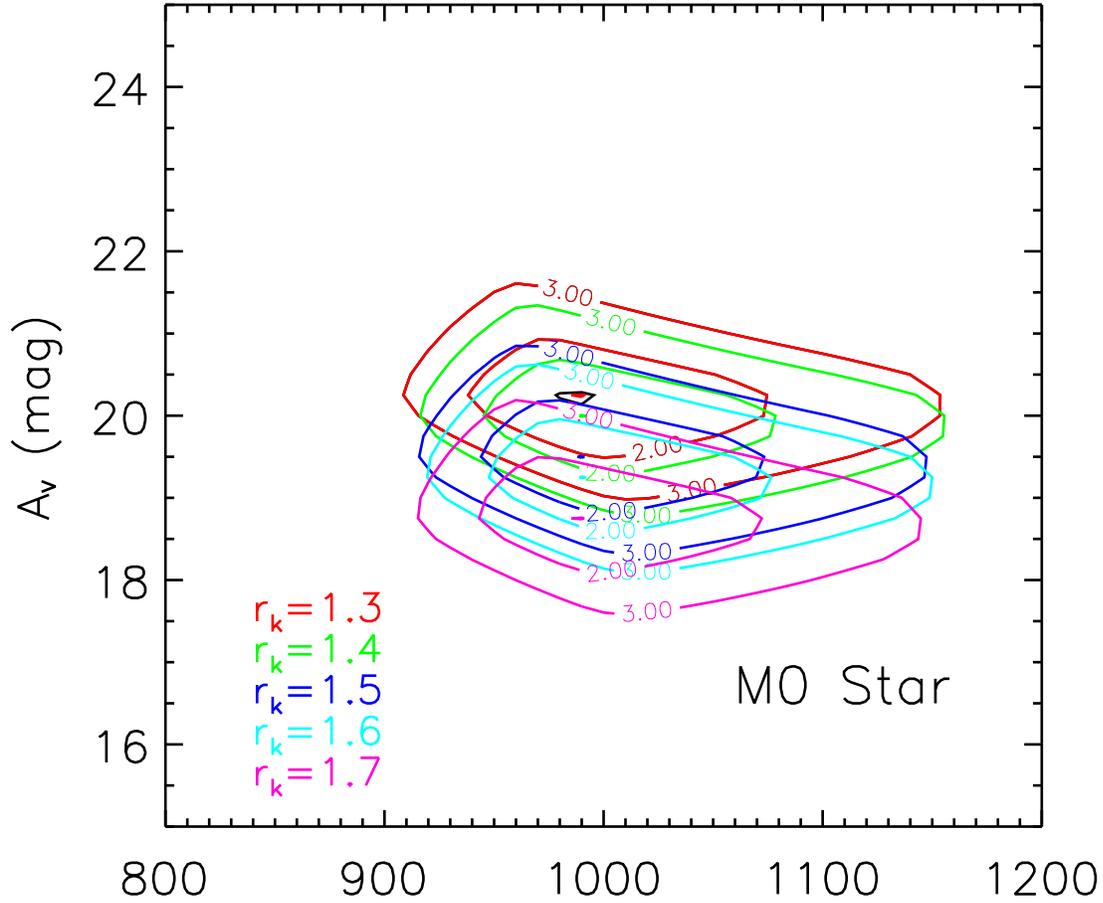} 
\caption{The $\chi^{2}$ surface plot for model fits to the V1647~Ori 1--4~$\mu$m spectrum.   In this plot we used an M0~V star as representative of V1647~Ori to plot the derived A$_V$ and T$_{dust}$ values for different r$_K$ values ranging from 1.3 to 1.7.  The surfaces for $\chi^{2}$ = 1.01, 2, and 3 are shown.  The 3.0  surface corresponds approximately to 1$\sigma$.  The range of acceptable values of A$_V$ and T$_{dust}$, $\Delta$A$_V$ and $\Delta$T$_{dust}$, are $\sim$2 magnitudes and $\sim$250~K, respectively.
\label{TB_Figure3}}

\end{figure*}

A $\chi^2$ minimization search was performed over the three parameters, A$_v$, T$_{dust}$ and $\tau_{ice}$ as in Beck (2007).  To obtain meaningful uncertainties on the derived values, we have generated a range of $\chi^2$ surfaces for A$_v$ and T$_{dust}$ values for the K5 to M2 range of stellar types which best fit the K-band spectrum (e.g. Fig.~\ref{TB_Figure1}). A $\chi^2$ minimization search was performed over the three parameters, A$_v$, T$_{dust}$ and $\tau_{ice}$.  As found by Beck (2007), the optical depth of the water-ice absorption feature did not affect the models appreciably in that only small changes in $\tau_{ice}$ are seen for a wide range of A$_v$ and T$_{dust}$ values.  For the analysis involving a range of stellar types, the infrared veiling value was adopted to be r$_k$=1.5 and the shape of the IR excess incorporated water vapor excess as in Fig.~\ref{TB_Figure2}.  Fig.~\ref{TB_Figure3} presents the $\chi^2$=1.01, 2.0 and 3.0 surfaces in A$_v$ and T$_{dust}$ for the models using K5~V (red), K7~V (green), M0~V (blue), M1~V (cyan), and M2~V (magenta) spectral template stars.  Because the models are fitting a broad spectral range, the 1$\sigma$ surface (approximated by $\chi^2$=3.0) is very narrow in A$_v$ and T$_{dust}$.  Best-fit A$_v$ values range from 17.5 to 21.5 magnitudes, while best-fit T$_{dust}$ values span the range 900~K to 1150~K.  Based on these results and the K-band fits, we adopt the M0~V stellar template star to be the most accurate model, with an uncertainty of $\pm$2 spectral classes.

\begin{figure*}[tb] 
\includegraphics*[angle=0,scale=1.0]{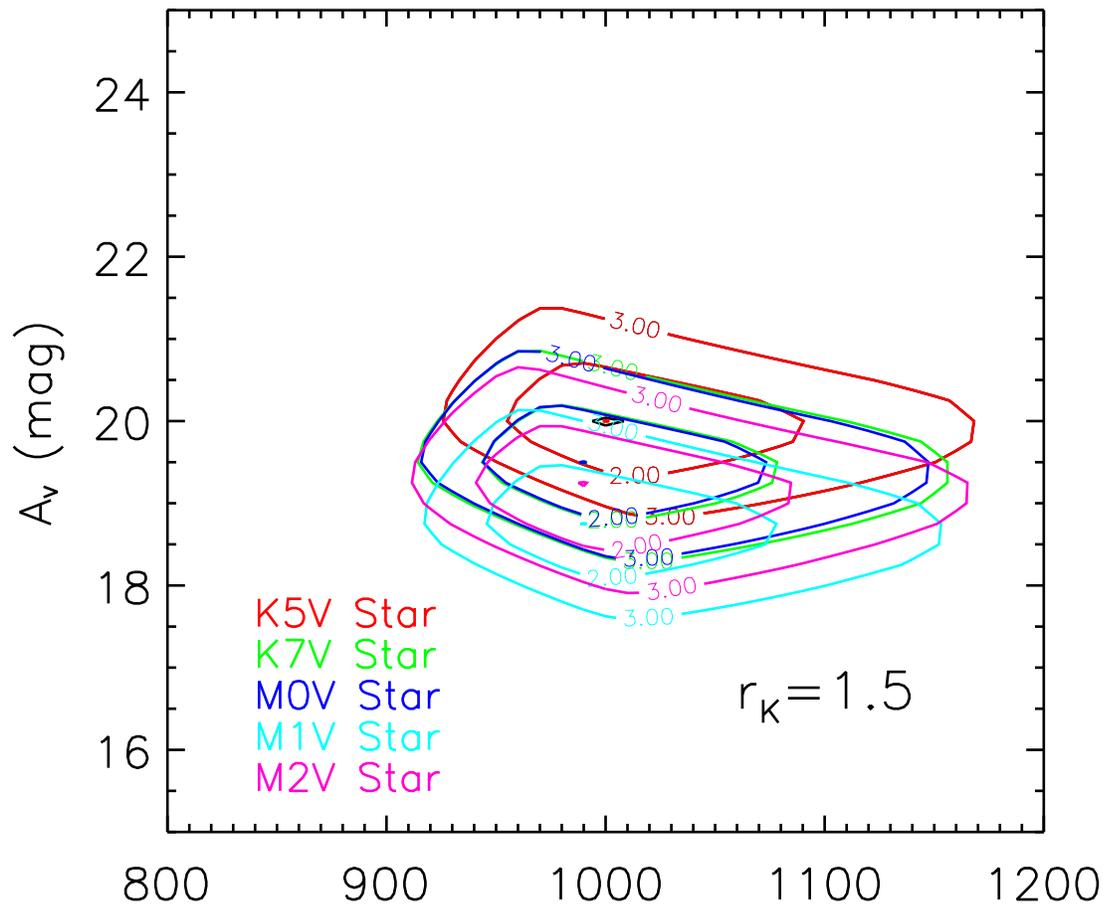} 
\caption{A similar plot to that shown in \ref{TB_Figure3} but for a range of spectral types, K5~V to M2~V for fixed r$_K$=1.5.  Again, the 1$\sigma$ surface corresponds to a range of $\Delta$A$_V\sim$2 and $\Delta$T$_{dust}\sim$250~K.   
\label{TB_Figure4}}

\end{figure*}

To further define the uncertainties in the spectral models, a second $\chi^2$ set of searches was performed over A$_v$ and T$_{dust}$.  For this search, a spectral type of M0~V was adopted and the infrared veiling value, r$_k$, was the free parameter.  Fig.~\ref{TB_Figure4} presents the $\chi^2$=1.01, 2.0 and 3.0 surfaces for A$_v$ vs. T$_{dust}$ using r$_k$=1.3 (red), 1.4 (green), 1.5 (blue), 1.6 (cyan) and 1.7 (magenta).  Again, for a range of infrared veiling values, models that fit the broad 1.0--4.1~$\mu$m spectral range show very small variation in the $\chi^2$ surface.  Based on these analyses, the best-fit model for the 1.0--4.1~$\mu$m near-IR spectra of V1647~Ori (including the effects of water vapor absorption) is adopted to be an M0$\pm$2 star with A$_v$=19$\pm$1 magnitude, r$_k$=1.5$\pm$0.2, and a T$_{dust}$=1000~K$\pm$100~K. 

\begin{figure*}[tb] 
\includegraphics*[angle=0,scale=0.9]{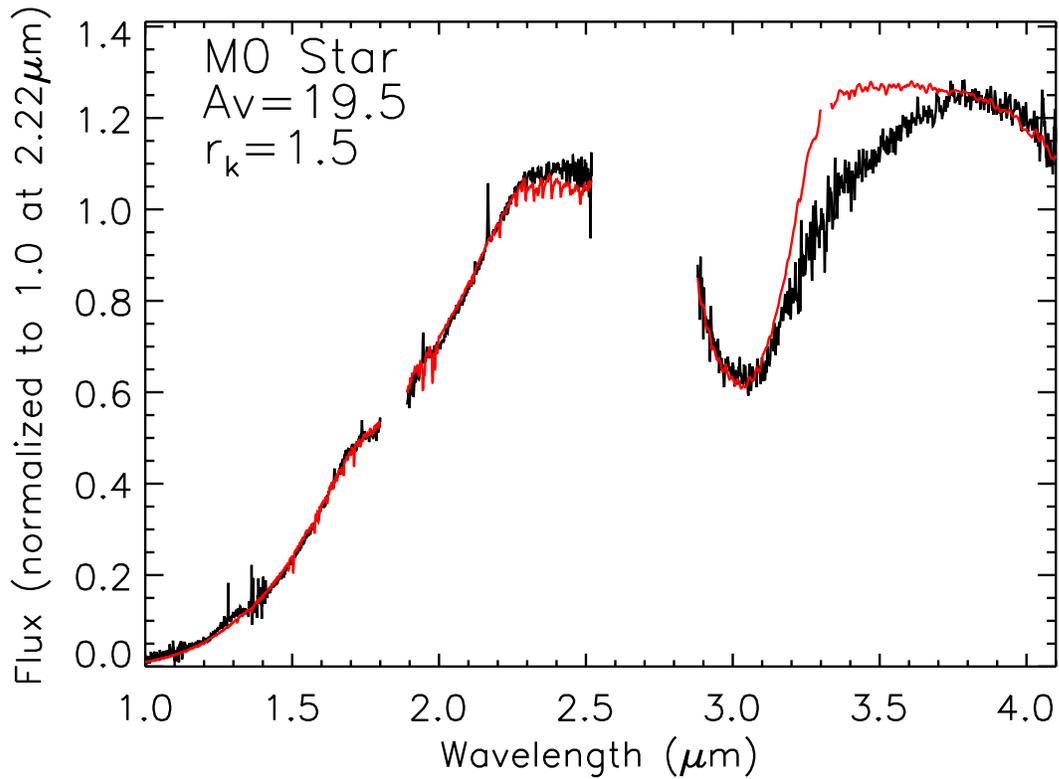} 
\caption{A plot of the observed (black) and best-fit model spectrum (red) using an M0~V spectral type template star, A$_V$=19, r$_K$=1.5, the water vapor model defined in the text (Fig.~\ref{TB_Figure2}), and incorporating a model ice absorption band from laboratory ices.  The poor fit from 3.2 to 3.8~$\mu$m is due to the model not including the long-wavelength wing of the 3.1~$\mu$m water ice absorption feature that has, as yet, an unknown cause. 
\label{TB_Figure5}}

\end{figure*}

\begin{figure*}[tb] 
\includegraphics*[angle=0,scale=1.0]{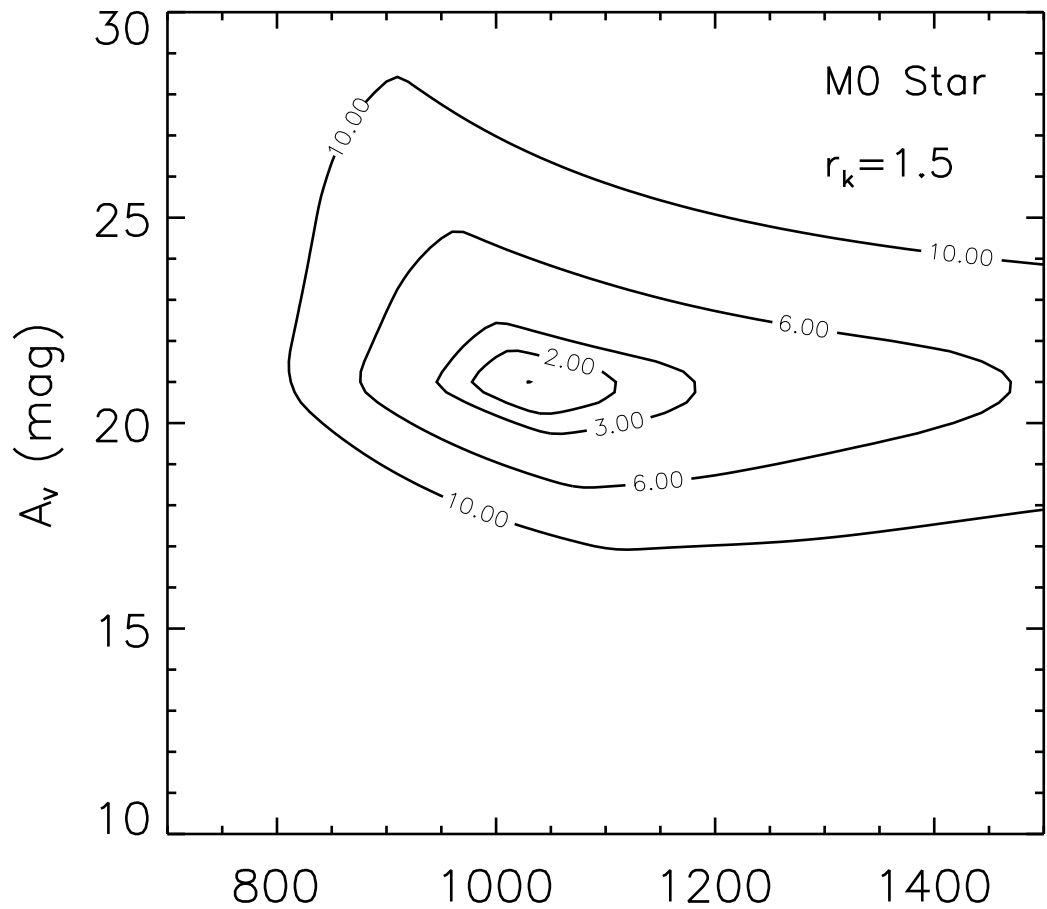} 
\caption{A $\chi^{2}$ surface plot for the final adopted values for the best-fit model.  Specifically, these are an M0~V star, r$_K$=1.5 and the water vapor model.  The $\chi^{2}$=10 surface corresponds approximately to 3$\sigma$.  
\label{TB_Figure6}}

\end{figure*}

Fig.~\ref{TB_Figure5} presents the full 1.0--4.1~$\mu$m spectrum of V1647~Ori with the best-fit model overplotted in red.  Unfortunately, the models cannot explain the known long-wavelength absorption wing to the 3.0~$\mu$m water-ice absorption feature (the 3.27--3.70~$\mu$m region was excluded from the spectral model fits -- Smith et al. 1989; Beck 2007).  Fig.~\ref{TB_Figure6} shows the full $\chi^2$=1.01, 2.0, 3.0, 6.0, and 10.0 surfaces for the model using an M0~V template star.  The $\chi^2$=10.0 surface is a measure of the 3$\sigma$ uncertainties in the model.  Visual extinctions as low as $\sim$11 magnitudes A$_v$ as measured by other studies (e.g. AP07) are excluded at a $>$3$\sigma$ level of confidence in our models.

\subsection{Additional determinations of visual extinction}
Of crucial importance to the determination of the physical parameters of V1647~Ori is the determination of reliable visual extinction values.  With the dramatic optical fading of V1647~Ori of over 5 magnitudes, A$_V$ values towards the young star likely vary with time as suggested by Reipurth \& Aspin (2004).  McGehee et al. (2004) determined that at optical wavelengths the change in color of the source was a result of {\it both} an A$_V$ change and the addition of significant blue accretion luminosity.  However, in the near-IR, the effect of such an increase in optical brightness would probably be much smaller.  Below, we examine our near-IR photometry and spectroscopy and attempt to derive A$_V$ values using several different and independent methods and compare them to A$_V\sim$19$\pm$1 derived from the spectral template fitting.

\subsubsection{A$_V$ from Near-IR colors}
From our calibrated near-IR spectrum we can obtain estimates of the broad-band J, H, and K magnitudes of V1647~Ori.  The determination of J and K magnitudes from this spectrum was performed using our ULBCam H band imaging magnitude of 11.96 to determine an absolute calibration.  Using the conversion defined by Tokunaga \& Vacca (2005) for J and K bands, we obtain J=14.8, H=11.96, and K=9.87 and hence J-H=2.8 and H-K=2.1. Dereddening these values along a reddening vector to the CTTS locus defined by Meyer et al. (1997) gives an A$_V\sim$17$\pm$4 magnitudes where the uncertainty is derived from the errors on the photometric values.

\begin{figure*}[tb] 
\epsscale{0.8}
\plotone{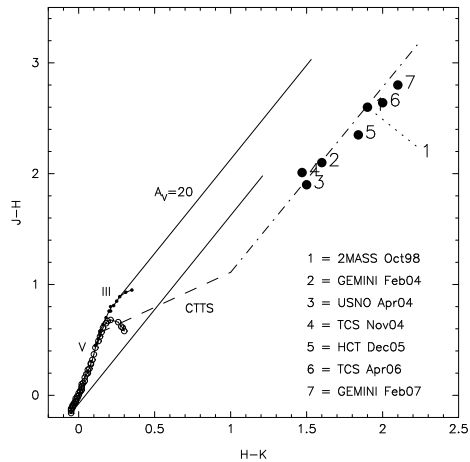} 
\caption{A near-IR J-H vs. H-K color-color diagram showing the published photometry for V1647~Ori from the pre-outburst and outburst phases together with our quiescent phase values.  The photometric values are shown as solid circles and the ID numbers refer to the source and time of the observations.  The legend shows the identification of the photometry, relating ID number to source of data and observing time.  Point 1 is from 2MASS data, point 2 from Reipurth \& Aspin (2004) Gemini data, point 3 from the United States Naval Observatory telescope as presented in McGehee et al. (2004), points 4 and 6 are from the Himalayan Chandra Telescope (HCT) presented in Ojha et al. (2006), point 5 is from the Telescopio Carlos Sanchez (TCS) at the Teide Observatory presented by Acosta-Pulido et al. (2007), and point 7 is from the data presented here. The colors of main sequence dwarfs (labeled V), giants (labeled III), and classical T ~Tauri stars (labeled CTTS) are also shown.  The CTTS locus (the dashed line) is from Meyer et al. (1997).  Reddening vectors corresponding to A$_V$=20 magnitudes are shown extending from the extremes of the main sequence/giant branch (solid lines) and the same reddening vector is shown extending from the extreme of the CTTS colors (dash-dotted line).  
\label{jhk-cc}}

\end{figure*}

\begin{figure*}[tb] 
\epsscale{0.9}
\plotone{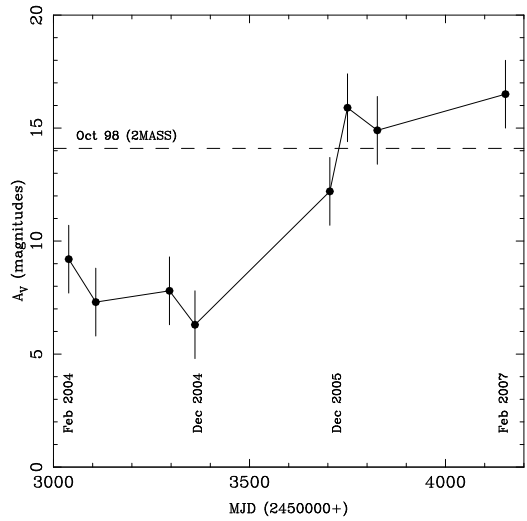} 
\caption{A plot of A$_V$ vs. Modified Julian Date (MJD) showing the change in A$_V$ with time from pre-outburst to outburst, and into quiescence.  The A$_V$ values were derived from the J-H and H-K colors in Fig.~\ref{jhk-cc} by dereddening onto the CTTS locus from Meyer et al. (1997).  The dashed line represents the A$_V$ value derived from the 2MASS J, H, and K' colors which we adopt as our pre-outburst value.  The outburst values are those from February 2004 to February 2006 while the quiescent value is from our February 2007 data.  Clearly, in early 2006 V1647~Ori reverted to its pre-outburst color. The difference in A$_V$ between the pre-outburst and outburst values is $\Delta$A$_V\sim$8.5 magnitudes. 
\label{jhk-av}}

\end{figure*}

An interesting feature of the published near-IR colors for V1647~Ori is that values from October 1998 (2MASS), March 2004 (Reipurth \& Aspin 2004), and May 2006 (AP07) all lie close to the same reddening vector and have a range of A$_V$ values from $\sim$9 (at the peak of the outburst phase) to $\sim$17 magnitudes (one year into the quiescent phase).  This trend demonstrates that the A$_V$ towards the star decreases during outburst and increase as the source fades toward quiescence.  Fig.~\ref{jhk-cc} shows this graphically where we plot published J-H and H-K colors covering the pre-outburst, outburst and the quiescent phase (our data).  Using these published near-IR colors, we have derived the A$_V$ values for each observation by dereddening onto the CTTS locus.  The estimated A$_V$ values are plotted against time in Fig.~\ref{jhk-av}.  It is clear from this plot that the A$_V$ towards V1647~Ori is considerably larger in its quiescent phase than during the outburst phase.   We have considered whether a change in T$_{dust}$ near the star could mimic the change in A$_V$ seen in Fig.~\ref{jhk-av}.  Near-IR J-H and H-K colors for pure Planck functions with temperatures in the range 950~K to 1350~K do give similar colors as V1647~Ori exhibits in Fig.~\ref{jhk-av}, however, for this to be the underlying cause of the observed color changes from outburst to quiescence we would have to be observing purely thermal emission from the cool black-body with an insignificant contribution from the stellar photosphere.  Since we detected photospheric features in the quiescent phase K-band spectrum of V1647~Ori, this interpretation is clearly invalid.

It also appears that {\it i)} the near-IR colors of V1647~Ori are little effected by the addition of the hot accretion component to the luminosity during the outburst phase, and {\it ii)} that the near-IR thermal excess remained approximately constant from the outburst to the quiescent phase with a change in visual extinction of $\Delta$A$_V\sim$8 magnitudes.  The only way {\it (i)} above can occur is if the accretion luminosity is 'gray' at near-IR wavelengths i.e. the accretion luminosity increases the emergent flux but does not change the observed near-IR colors of the source.  Gray absorption has been previously observed in accreting young stars in Orion~A by Carpenter, Hillenbrand, \& Skrutskie (2001) and interpreted as being due to large particles. However, whether a similar mechanism can be invoked for emission is unclear.   As to {\it (ii)} above, it would seem that the level of accretion in the pre-outburst phase (when the near-IR colors were measured by 2MASS i.e. October 1998) must have been similar to that in quiescence (in February 2007), otherwise the SEDs would not be close to identical. 

\subsubsection{A$_V$ from near-IR [Fe~II] line ratios}
Reipurth et al. (2000) used the two near-IR [Fe~II] lines at 1.275 and 1.644~$\mu$m to obtain an estimate of the visual extinction towards the [Fe~II] emitting region in a number of HH energy sources.  This is possible since the transitions forming the lines arise from the same upper level and therefore the line ratio is fixed.  We note that the estimate obtained will be the line-of-sight extinction to the [Fe~II] emitting region which may be located some distance from the stellar photosphere and possibly extended in nature.  Nonetheless, using the formula shown in Reipurth et al. (2000), inverted to estimate A$_V$ i.e.

\begin{equation}
A_V = -25 \times log_{10}((I_{1.275}/I_{1.644})/1.36)
\end{equation}

for an extinction law that gives A$_V\sim$10$\times$E$_{J-H}$, the [Fe~II] line fluxes in Table~2 give a ratio I$_{1.275}$/I$_{1.644}$=0.3 and therefore an A$_V\sim$16$\pm$3 magnitudes.  The error is estimated from repeat measurement of the (weak) line fluxes.

\subsubsection{A$_V$ from the 3~$\mu$m water ice absorption feature}
As outlined in Beck (2007), deriving A$_V$ values from the 3~$\mu$m water ice absorption feature is somewhat unreliable for A$_V>$20 magnitudes due to uncertainties in the relationship between ice band optical depth and A$_V$.  Also, grains nearer to the star that still contribute to the line-of-sight extinction, may not have ice mantles if the local temperature is $>$200~K.  Nonetheless, since all indications are that for V1647~Ori we are dealing with A$_V$ values $<$20 magnitudes (e.g. AP07), we proceed with the derivation. 

Following Beck (2007) and using the relation presented in Teixeira \& Emerson (1999) between water column density, N(H$_2$O), and A$_V$, for a peak optical depth in the water band $\tau_{peak}\sim$0.7, and a FWHM of the band of 320~cm$^{-1}$, we obtain N(H$_2$O)$\sim$1.0~10$^{18}$~cm$^{-2}$.  This value is almost identical to that derived by Rettig et al. (2006) i.e. 1.2~10$^{18}$~cm$^{-2}$.  From Fig.~5a in Teixeira \& Emerson (1999), we see that the above column density corresponds to an A$_V\sim$13$\pm$4 magnitudes.  However, as found in Beck (2007), changing the value of A$_V$ used resulted in only small changes in the model continuum level and did not substantially change the derived ice absorption optical depth.  In light of this, we consider that the A$_V$ derived using the ice absorption optical depth is likely not as accurate as the values from the other determinations above. 

\subsubsection{A$_V$ from near-IR H~I line ratios}
Another method for determining A$_V$ towards the emitting region is the measured ratio of the H~I lines Br$\gamma$ (2.166~$\mu$m) and Br$\alpha$ (4.052~$\mu$m).  This determination requires the assumption that only Ly$\alpha$ photons are optically thick (i.e. Case~B of Baker \& Menzel 1938) over a wide range of temperatures and densities. If that assumption is correct then the ratio Br$\alpha$/Br$\gamma$ should be in the range 2.85--3.17 (Hummer \& Storey 1987).  Due to the relatively low signal to noise of Br$\alpha$ it is difficult to estimate the line flux accurately.  Our best value of the above ratio, obtained from repeated measurements of the Br$\alpha$ line flux, is 2.0$\pm$0.6.  This suggests that the Br$\alpha$ line flux is less that expected from Case~B and, hence, the Case~B assumption is likely invalid.  As a result of this, we do not attempt to derive A$_V$ values from the H~I emission ratios.

\subsubsection{Summary of derived A$_V$ values}
To summarize, our four independent determinations of A$_V$ i.e. from spectral template fitting (19$\pm$1), near-IR colors (17$\pm$4), [Fe~II] line ratios (16$\pm$3), and 3~$\mu$m ice absorption optical depth (13$\pm$4), all give reasonably consistent results.  However, since we put lower weight on the value determined from the 3~$\mu$m ice absorption feature, and are most confident in the A$_V$ value derived from our spectral template fitting, we adopt A$_V\sim$19$\pm$2 magnitudes as the visual extinction towards V1647~Ori in February 2007 and note that the A$_V$ values derived from near-IR colors and [Fe~II] line ratios are consistent with this value (within the associated uncertainties).  Additionally we have found that there was a systematic decrease in A$_V$ towards V1647~Ori during the outburst phase.  This was mirrored by an increase in A$_V$ as the source faded to, and into, quiescence. 

\subsection{Derived physical parameters}

\subsubsection{Luminosity and absolute magnitude}
Above, we determined that V1647~Ori is likely a star of spectral type M0$\pm$2 with a T$_{eff}$ of $\sim$3800~K, a visual extinction A$_V\sim$19$\pm$2 magnitudes (in February 2007), a veiling of r$_K$=1.5$\pm$0.2, and an apparent K magnitude of m$_K\sim$9.90 (in February 2007).  Using the bolometric correction (M$_{bol}$--M$_{v}$=-1.3) and M0 color (V--K=3.7) from Table~A5 in Kenyon \& Hartmann (1995) and the equations for M$_{bol}$ and L$_{*}$ from Greene \& Lada (1997), we estimate that V1647~Ori has an absolute bolometric magnitude of M$_{bol}\sim$2.9$\pm$0.4 and a luminosity of L$_{*}\sim$5.2$\pm$2~L$_{\odot}$ for a distance of d=450~pc (m-M=8.3).  The error on our value of L$_{*}$ takes into account the uncertainties on r$_K$, A$_V$, and the spectral type quoted above, but not uncertainties in distance.  Thus, we estimate that the star has a radius, R$_{*}$, of $\sim$5~R$_{\odot}$ using the standard T$_{eff}$ vs. L$_{*}$ relationship.

\subsubsection{Stellar mass and age}
\begin{figure*}[tb] 
\includegraphics*[angle=90,scale=0.7]{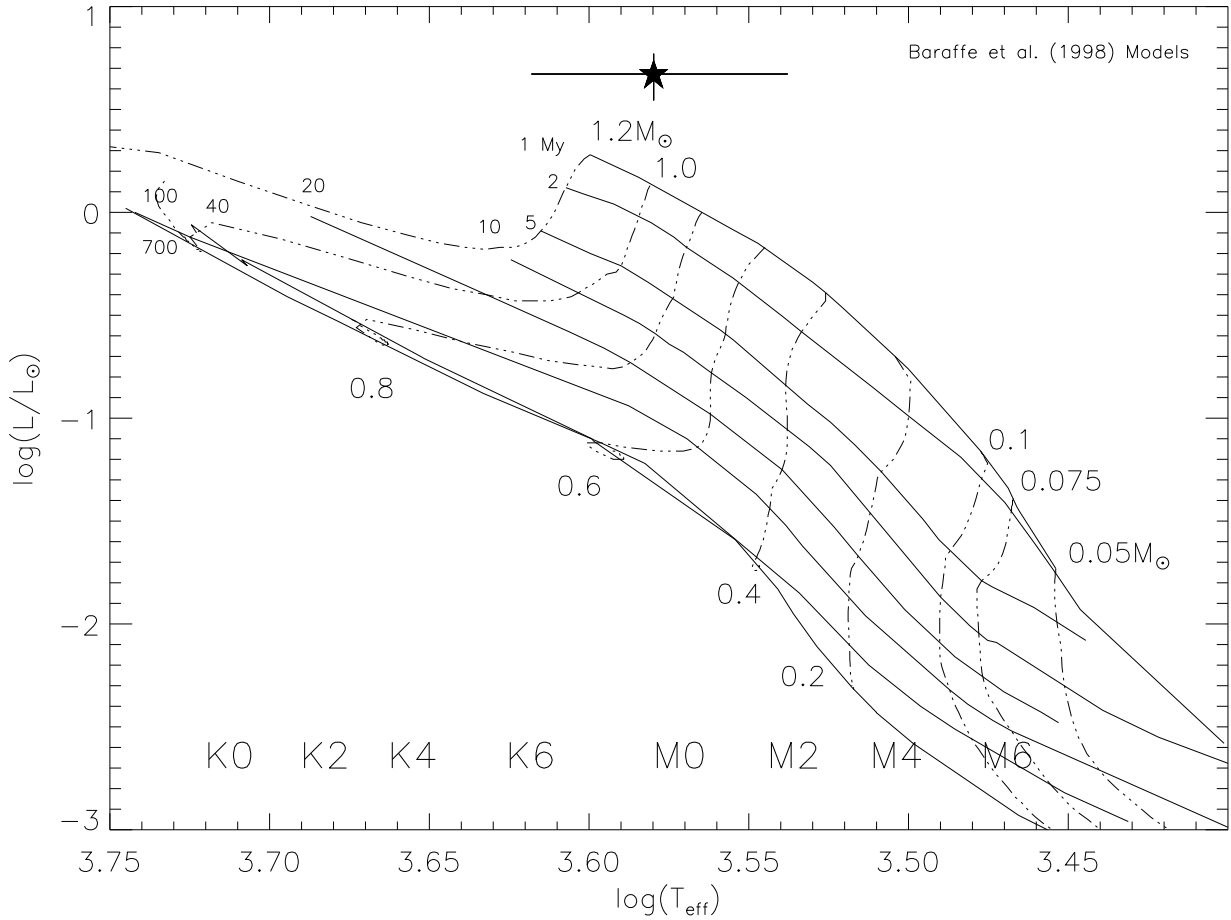} 
\caption{A comparison of our derived stellar luminosity L$_{*}$=5.2~L${\odot}$ and effective temperature  T$_{eff}$=3800~K with evolutionary track models from Baraffe et al. (1998).  The location of V1647~Ori is indicated by a filled star with associated uncertainties displayed.   The location of V1647~Ori in this plot suggests it is a star of mass $\sim$1~M$_{\odot}$ with an age younger than $\lesssim$0.5~Myr. 
\label{baraffeHR}}

\end{figure*}

\begin{figure*}[tb] 
\includegraphics*[angle=90,scale=0.7]{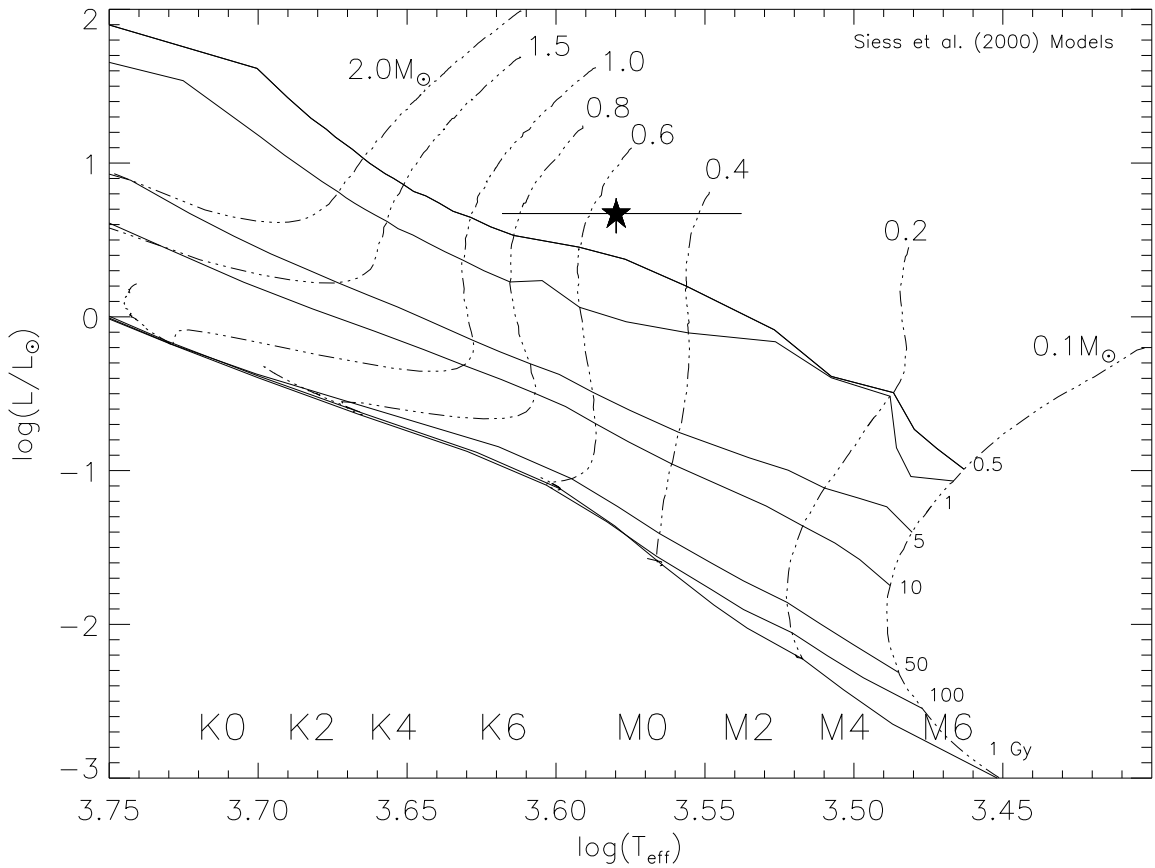} 
\caption{A similar plot to that shown in Fig.~\ref{baraffeHR} but for the evolutionary tracks models of Siess et al. (2000).   Again, V1647~Ori is indicated by the filled star with associated uncertainties displayed.  The location of V1647~Ori in this plot suggests a stellar mass of ~$\sim$0.55~M$_{\odot}$ and an age of $<$0.5~Myrs.
\label{siessHR}}

\end{figure*}

Using the stellar luminosity and spectral type derived above, we can place V1647~Ori on pre-main sequence stellar tracks using results from numerical models.   Our adopted values are L$_{*}\sim$5.2$\pm$2~L$_{\odot}$ and T$_{eff}\sim$3800$\pm$350~K.  The temperature (and associated uncertainty) comes from our adopted spectral type of M0 ($\pm$2 sub-classes) using the values in Table~A5 of Kenyon \& Hartmann (1995).   We have placed V1647~Ori on the evolutionary tracks of both Baraffe et al. (1998) and Siess et al. (2000).   Figs.~\ref{baraffeHR} and \ref{siessHR} show its location on these model tracks.  The models give masses and ages for V1647~Ori of 1~M$_{\odot}$/$<$0.5~Myrs, and 0.55~M$_{\odot}$/$<$0.5~Myrs, respectively.  We conclude therefore that, conservatively, V1647~Ori is probably of mass 0.8$\pm$0.2~M$_{\odot}$ with an age of $<$~0.5~Myrs.  

\subsubsection{Mass accretion rate and accretion luminosity}
It is possible to determine mass accretion rates for young stars like V1647~Ori using H~I emission line strengths.   Following Muzerolle, Hartmann, \& Calvet (1998), we can use the dereddened flux of the near-IR  emission lines Pa$\beta$ and Br$\gamma$ to derive accretion luminosities, L$_{acc}$, and following Gullbring et al. (1998) determine a mass accretion rate, \.M.  The reader is referred to Eqns.~1 \& 2 in Muzerolle, Hartmann, \& Calvet (1998) and Eqn.~8 in Gullbring et al. (1998) for further details of the derivation.  For the calculations below, we assume M$_{star}$=0.8$\pm$0.2~M$_{\odot}$, L$_{*}$=5.2$\pm$2~L$_{\odot}$, T$_{star}$=3800~K, R$_{in}$=5~R$_{star}$, A$_V$=19, the extinction law shown in Eqn.~1 above, and d=450~pc. R$_{in}$ is the inner radius of the accretion disk. 

Dereddened fluxes and derived A$_\lambda$ values for all detected optical and near-IR emission lines are shown in Table~2.  Using the dereddened Pa$\beta$ flux, we obtain L$_{acc}\sim$3.5~L$_{\odot}$ and log(\.M)$\sim$--6.1 or \.M$\sim$8.8~10$^{-7}$~M$_{\odot}$~yr$^{-1}$.  With the uncertainties associated with the parameters in the formula used, the range of possible values for  L$_{acc}$ are 2.1--5.5~L$_{\odot}$ and for \.M is 5.4~10$^{-7}$ to 1.4~10$^{-6}$ ~M$_{\odot}$~yr$^{-1}$.  Similarly, for the dereddened Br$\gamma$ flux, we obtain L$_{acc}\sim$4.6~L$_{\odot}$ and a log(\.M)$\sim$--5.92 or \.M$\sim$1.2~10$^{-6}$~M$_{\odot}$~yr$^{-1}$.  Considering the quoted uncertainties, we obtain a range of possible values of L$_{acc}$ of 2.8--7.7~L$_{\odot}$ and \.M=7.2~10$^{-7}$ to 2.0~10$^{-6}$~M$_{\odot}$~yr$^{-1}$.  

Another method to derive \.M was discussed by White \& Basri (2003) and Natta et al. (2004).  This derivation relates the width of the optical H$\alpha$ emission line to accretion rate.  The quantity used for this is the H$\alpha$ full width at 10\% peak intensity and an approximately linear relationship was found between this and log(\.M) by Natta et al. (2004).  Even though the H$\alpha$ line profile appears asymmetric we apply this technique to investigate its applicability.   Using Eqn.~1 from White \& Basri (2003), and a measured H$\alpha$ at 10\% width of 740~km~s$^{-1}$, we obtain a log(\.M)$\sim$--5.7 or \.M$\sim$2~10$^{-6}$~M$_{\odot}$~yr$^{-1}$.  This derivation assumes that the H$\alpha$ emission is solely the result of the accretion processes and not contaminated by stellar winds (absorption/emission) and/or HH flows/jets (emission) which, due to the presence of an asymmetric profile, may not be the case.  However, we note that if we double the blue-wing 10\% emission width to compensate for the red-shifted absorption, we would obtain a larger value of H$\alpha$ at 10\% and hence, a larger accretion rate.   

To summarize, our two estimates of accretion luminosity and rate give reasonably consistent results and we adopt values of L$_{acc}\sim$4$\pm$2~L$_{\odot}$ and \.M$\sim$1.0$\pm$0.5~10$^{-6}$~M$_{\odot}$~yr$^{-1}$, respectively.

\subsubsection{Bolometric Luminosity}
Our estimate of L$_{bol}$ is the sum of our derived L$_{*}$ and L$_{acc}$ values, or, L$_{bol}$=9.2$\pm$3~L$_{\odot}$.  Andrew, Rothberg, \& Simon (2004) calculated an outburst luminosity of L$_{bol}\sim$3.5~L$_{\odot}$ and noted it was similar to the one derived by Lis, Menten, \& Zylka (1999) once a distance correction was applied (they assumed d=400~pc). They also considered their luminosity estimate a lower limit due to the lack of information on the spectral energy distribution (SED) peak in the far-IR and the linear interpolation used to define the SED.  We note that the L$_{bol}$ estimate from Lis, Menten, \& Zylka (1999) only considered emission from two modified Planck functions that fitted the IRAS and sub-mm fluxes and did not include a stellar component.  In addition, it is clear that the IRAS fluxes can contain contributions from nebular emission and from other stars in the beam thus contaminating the measurements.  We also note that Andrews, Rothberg, \& Simon (2004) did not quote any formal uncertainty on their L$_{bol}$ determination.  The error value is probably significant in light of the points related above.  We consider our estimate of L$_{bol}$ to be robust in that it has considered both stellar luminosity and accretion luminosity independent of the source SED.   Since the SED of V1647~Ori is very similar pre-outburst and in the quiescent phase (see Section 3.7.5), we consider that the L$_{bol}$ of the source during both these periods is 9.2$\pm$3~L$_{\odot}$.

\subsubsection{Spectral Energy Distribution}
\begin{figure*}[tb] 
\includegraphics*[angle=270,scale=0.7]{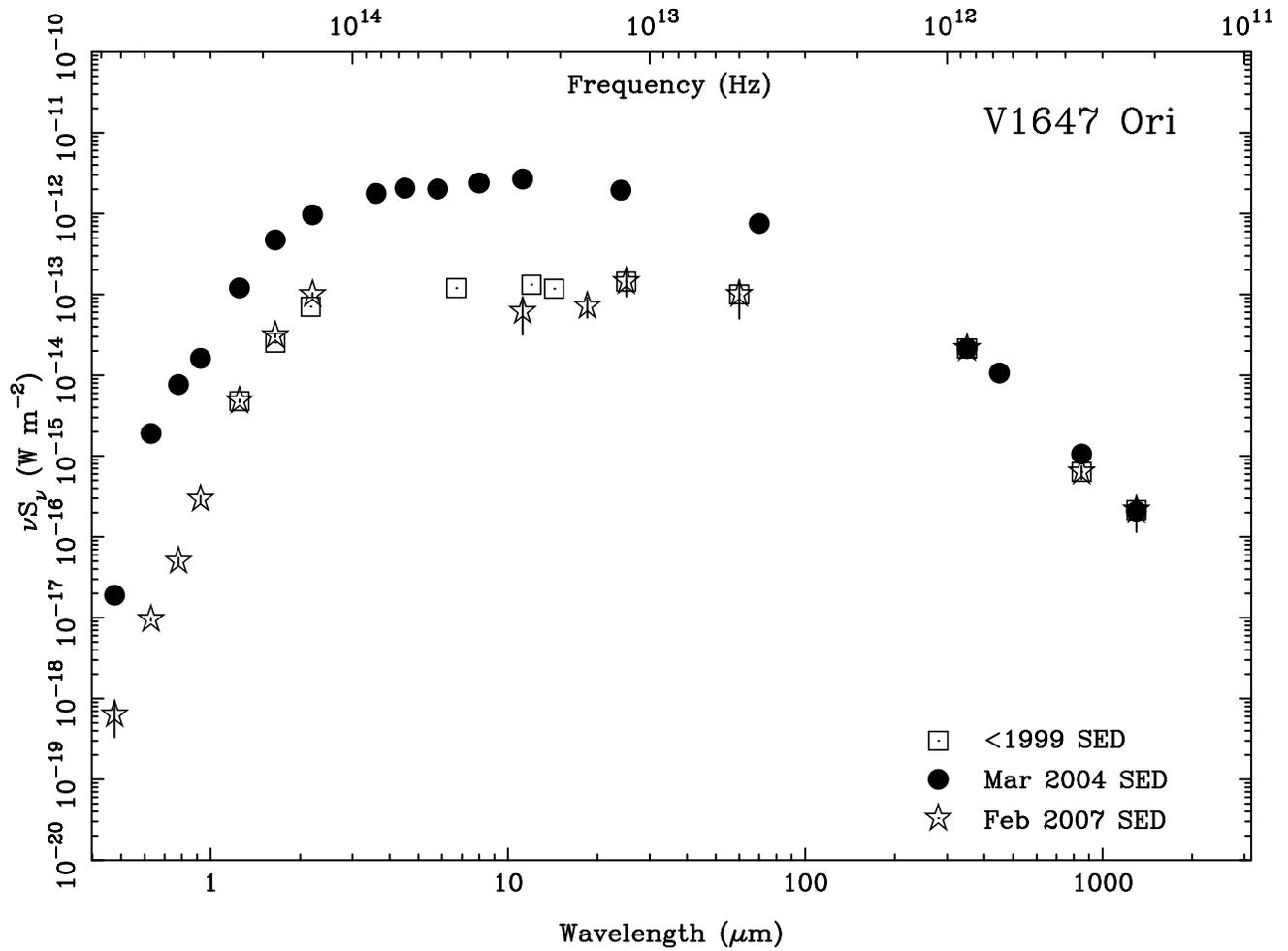} 
\caption{The spectral energy distribution (SED) of V1647~Ori.  Plotted are the pre-outburst SED (open squares) and the outburst SED (filled dots) both from Andrews, Rothberg, \& Simon (2004), and the quiescent phase SED from our data (open stars).  Uncertainties on the quiescent phase data are indicated.  Note that the pre-outburst and quiescent phase SEDs are very similar. 
\label{sed}}

\end{figure*}

The SED of V1647~Ori is shown in Fig.~\ref{sed}.  The open stars are the data from February 2007 while the open boxes and solid points are the pre-outburst and outburst SEDs, respectively.  We have used the same data as Andrews, Rothberg, \& Simon (2004) longward of 25~$\mu$m (i.e. IRAS, and JCMT data) and note that they state that the pre-outburst and outburst photometry from 350--1300~$\mu$m was unchanged by the eruption.  It is obvious that the quiescent phase SED is almost identical in shape and flux level to the pre-outburst SED.  This suggests that the state of V1647~Ori in February 2007 was very similar to that in the pre-outburst period.     

\subsection{The Mid-IR Characteristics}
\begin{figure*}[tb] 
\epsscale{1.0}
\plotone{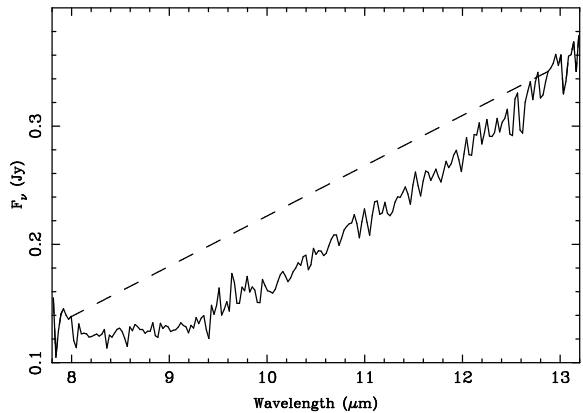} 
\caption{The 8--13~$\mu$m spectrum of V1647~Ori taken with the Gemini Michelle imaging spectrometer on UT April 5, 2007.  The spectrum was ratioed with a telluric standard star and calibrated using the Michelle photometry shown in Table~1.  Hence, the flux at 11.2~$\mu$m is 0.23~Jy.  The dashed line is a linear interpolation between 8 and 13~$\mu$m (see text for details).
\label{mirspecplot}}

\end{figure*}

In  Fig.~\ref{mirspecplot} we show the mid-IR N-band spectrum of V1647~Ori from UT 5 April 2007.  These data were calibrated using the mid-IR photometric data presented in Table~2.  The spectrum is featureless apart from a change in slope around $\sim$9.5~$\mu$m.  Shortward of this wavelength the spectrum is relatively flat.  At $\sim$9.5~$\mu$m, however, the slope increases rapidly to the edge of the band.  A previous mid-IR spectrum from Andrews, Rothberg, \& Simon (2004) taken in March 2004 did not show this change in spectral index.   Their spectrum rose to the red with a power-law fit F$_{\nu}\sim\lambda^{2.5}$.  A similar fit to the 9.2--13~$\mu$m region of our spectrum gives F$_{\nu}\sim\lambda^{2.8}$. 

One difference between our N-band spectrum and that of Andrews et al. is that the flux level has decreased significantly from $\sim$7~Jy to $\sim$0.14~Jy at 10~$\mu$m, a decline of a factor $\times$50.  In a study of the pre-outburst infrared characteristics of V1647~Ori, \'Abr\'aham et al. (2004) presented {\it ISOCAM} photometry at 6.7 and 14.3~$\mu$m.  If we interpolate these values to 10~$\mu$m we obtain a pre-outburst brightness of $\sim$0.4~Jy.  In addition, the {\it IRAS} 12~$\mu$m point-source flux for IRAS~05426-0007 is 0.53~Jy.  Given the uncertainties in the nature of the emission included in both the {\it ISOCAM} and {\it IRAS} beams, we consider this a reasonable agreement.

\begin{figure*}[tb] 
\epsscale{1.0}
\plotone{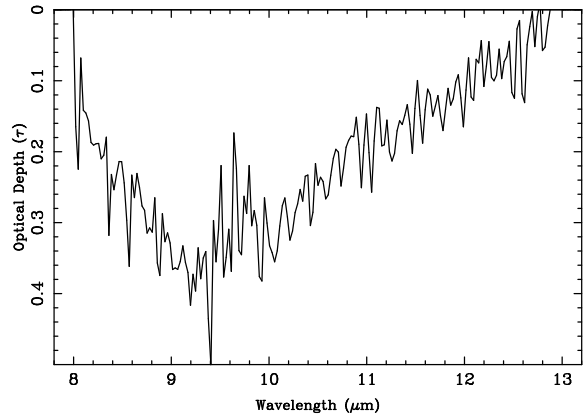} 
\caption{A plot of the 8--13~$\mu$m silicate absorption band optical depth extracted from the observed spectrum shown in Fig.~\ref{mirspecplot} by subtracting the observed the spectrum of V1647~Ori from the UT April 5, 2007 data.  The silicate absorption flux at each wavelength was subtracted from the flux from the linear interpolation as in Quanz et al. (2007).  The resultant profile was then converted to optical depth, $\tau$.
\label{tauplot}}

\end{figure*}

We consider that the aforementioned change in spectral slope at $\sim$9.5~$\mu$m is in fact indicative of the presence of weak silicate absorption through the passband.   We follow the analysis of Quanz et al. (2007), who studied a number of FUors in the mid-IR using {\it ISO} and IRS spectroscopic data.  They fitted a linear continuum between 8~$\mu$m and 13~$\mu$m and subtracted the observed spectrum from the result of the fitting procedure.   In Fig.~\ref{mirspecplot} we show such a linear fit as a dashed line.   Subtracting a dereddened (with A$_V$=19) observed spectrum from this fit and converting the result from flux to optical depth, $\tau$, we obtain the plot shown in Fig.~\ref{tauplot} where a relatively shallow yet typical silicate absorption band seen with a maximum optical depth of $\tau\sim$0.35.   Converting this value to an extinction at the absorption minimum i.e. 9.2~$\mu$m (we termed this A$_{9.2}$) results in a value of 0.38$\pm$0.15 magnitudes.  For comparison, the expected extinction at 9.2~$\mu$m for an A$_V$=19 is A$_{9.2}\sim$0.2.  Additionally, comparing our silicate absorption plot with those from Quanz et al. (2007, their Fig.~15) we see a number of FUors with similarly shaped band, specifically, V346~Nor and Z~CMa, which also have similar maximum optical depth to V1647~Ori (of $\tau\sim$0.3).   A cursory comparison of the silicate absorption band shape with the model silicate bands from Quanz et al. (2007, their Fig.~16) suggests that the V1647~Ori silicate band is most similar to that modeled using astronomical silicates as defined by Weingartner \& Draine (2001) and Draine (2003). 

In the observations of Andrews, Rothberg, \& Simon (2004) from UT March 11 2004, there is no evidence for any silicate feature be it in absorption or emission.  The mid-IR spectra of Quanz et al. (2007), from three epochs (UT October 20 2004, UT March 11 2005, and UT March 24 2005) were all found to show weak 10~$\mu$m silicate emission features.  Mosoni et al. (2005) shows an mid-IR spectrum of V1647~Ori taken UT March 1 2005.  This showed a similar weak silicate emission feature seen by Quanz et al.  Fedele et al. (2007a) also presented 10~$\mu$m spectra from three epochs (UT March 8 2004, UT February 21 2005, and UT January 10 2006).  In these, we believe that silicate emission is weakly present in only the February 2005 spectrum.  The progression therefore seems to be that {\it i)} silicate emission developed from a featureless red continuum between UT March 11 2004 and UT October 20 2004, {\it ii)} the emission persisted until at least UT March 24 2005, and {\it iii)} by UT January 10 2006 there was perhaps weak silicate absorption which grew a little deeper as time progressed to the level shown in our 10~$\mu$m spectrum from UT April 05 2007.  This result implies that there must have been significant dust evolution in V1647~Ori over the outburst period from March 2004 to April 2007.  

It is well known that some young eruptive variable stars show a 10~$\mu$m silicate feature in emission e.g. the FUors FU~Ori, BBW~76, and V1057~Cyg (Green et al. 2006).  Others, however, show a 10~$\mu$m silicate feature in absorption e.g. the FUors V346~Nor, Z~CMa, L1551~IRS5 (Green et al. 2006).  Silicate absorption is clearly produced when dense, cool silicate dust is located between the radiation source and the observer.  Silicate emission occurs when heated, optically thin, silicate dust, at a higher temperature than the source of the 10~$\mu$m continuum emission, is directly visible to the observer.  In addition to V1647~Ori, Quanz et al. (2007), studied of a number of FUors, using {\it Spitzer} and {\it ISO} data, and suggested that two categories exist, one with absorption created in a dense, dusty, and icy envelope (Category~1, herein C1), and the other with emission from the heated surface layers of a circumstellar/accretion disk (Category~2, herein C2). They considered the progression from C1 to C2 evolutionary in nature and related to the size of the cold, icy dust envelope.  Hence, C1 FUors would be younger objects that still possess extensive, dense, cold envelopes through which we are observing the FUor. C2 FUors would be more evolved in nature with only remnant envelopes and a more exposed accretion disk.  Since FUor eruptions are thought to repeat many times through the Class~I to Class~II stages, this is clearly a possibility although system orientation with respect to the line-of-sight may be important.  Even though V1647~Ori is not considered a FUor, perhaps the clearing/sublimation of dust towards the star resulted in the accretion disk being revealed (C2) and as the eruption and wind subsided significant dust re-accumulated/condensed along the line-of-sight, obscuring the accretion disk, and creating enough column density to create the weak silicate absorption feature (C1).

We defer further consideration of the 10~$\mu$m spectra to a future paper in which we will consider dust evolution more comprehensively using all published and archival 10~$\mu$m spectra together with new 10~$\mu$m spectra taken during our two year Gemini monitoring campaign (Aspin et al., in prep).  We conclude by noting that{\it i)} the appearance (and disappearance) of silicate emission supports the interpretation that the mid-IR flux originates from the circumstellar disk as suggested by Muzerolle et al. (2005) and \'Abr\'aham et al. (2006), and {\it ii)} the appearance of silicate absorption suggests that the line-of-sight column density of cold dust increased as the outburst faded and the source returned to its quiescent state.

\section{CONCLUSIONS}

From our optical, near-IR, and mid-IR imaging and spectroscopy of V1647~Ori dating from February to April 2007, approximately one year after its return to its pre-outburst optical brightness, we can conclude that:

\begin{itemize}
\item The associated nebula, McNeil's Nebula, remained faintly visible suggesting that quasi-static nebula material is still being illuminated by V1647~Ori.  

\item We confirm the findings of Fedele et al. (2007a) in that signposts of shock-excited emission, specifically, in the emission lines of [S~II], [Fe~II], and H$_2$ are present.  We consider that it is a distinct possibility that these emission lines result from a new Herbig-Haro flow. 

\item Our near-IR spectrum shows, for the first time, molecular overtone absorption from CO and atomic absorption from neutral Na and Ca.  We interpret this as evidence that we are now observing the stellar photosphere of V1647~Ori.  
\item We have modeled the near-IR spectrum using a template stellar spectrum including both water vapor and water ice absorption and have determined a best-fit parameter set of spectral type M0$\pm$2 sub-classes, A$_V$=19$\pm$2 magnitudes, r$_K$=1.5$\pm$0.2.

\item We derive values of M$_{bol}$=2.9$\pm$0.4 magnitudes and  L$_{*}$=5.2$\pm$2~L$_{\odot}$ for V1647~Ori. From comparison with theoretical evolutionary tracks, find that, for the adopted T$_{eff}\sim$3800~K, the star has a mass and age of 0.8$\pm$0.2~M$_{\odot}$ and $\lesssim$0.5~Myrs, respectively. 

\item From near-IR H~I line flux and the A$_V$ values derived above, we estimate the accretion luminosity and mass accretion rate in February 2007 was L$_{acc}$=4.0$\pm$2~L$_{\odot}$ and \.M=1.0$\times$10$^{-6}$~M$_{\odot}$~yr$^{-1}$, respectively.  This implies that V1647~Ori is still actively accreting circumstellar material even though it is almost optically invisible and that the accretion rate during the outburst must have been considerably larger that this value.

\item V1647~Ori is found to have a quiescent phase L$_{bol}$ of  9.25$\pm$3~L$_{\odot}$.

\item For the first time, we see evidence for silicate dust evolution in the mid-IR spectrum of V1647~Ori over the outburst to quiescence period.  We now observe weak silicate absorption at 10~$\mu$m whereas previously the silicate band was either absent or weakly in emission.   

\item Finally we note that, in February 2007, the spectral energy distribution, SED, of V1647~Ori appears remarkably similar to its pre-outburst SED suggesting that perhaps the derived accretion rate is the normal quiescent phase accretion rate for this object.
\end{itemize}

\vspace{0.3cm}

{\bf Acknowledgments} 

Based on observations obtained at the Gemini Observatory (under program identification GN-2007A-Q-33 and GS-2005B-Q-13), which is operated by the Association of Universities for Research in Astronomy, Inc., under a cooperative agreement with the NSF on behalf of the Gemini partnership: the National Science Foundation (United States), the Particle Physics and Astronomy Research Council (United Kingdom), the National Research Council (Canada), CONICYT (Chile), the Australian Research Council (Australia), CNPq (Brazil) and CONICET (Argentina).

Colin Aspin and Tracy Beck were visiting astronomers at the Infrared Telescope Facility, which is operated by the University of Hawaii under Cooperative Agreement no. NCC 5-538 with the National Aeronautics and Space Administration, Science Mission Directorate, Planetary Astronomy Program.

This material is based upon work supported by the National Aeronautics and Space Administration through the NASA Astrobiology Institute under Cooperative Agreement No. NNA04CC08A issued through the Office of Space Science.


\clearpage 

\begin{center}
\begin{deluxetable}{ccllccc}
\tablecaption{V1647~Ori Observing Log\label{obslog}}
\tablewidth{0pc}
\rotate
\tablehead{
\colhead{UT Date} & 
\colhead{MJD\tablenotemark{a}} & 
\colhead{Telescope/Inst.} & 
\colhead{Filters/Grism/$\lambda _C$/Slit Width} & 
\colhead{Exposure Times} & \colhead{Seeing} &
\colhead{Photometry}}

\startdata

20051010/11 & 3654/5 & Gemini-S/GMOS-S & [S~II] & 1200~s & 0.65$''$ & -- \\
20070213 & 4144 & UH2.2/ULBCam & H & 360~s & 0.80$''$ & 11.96$^m\pm$0.1$^m$ \\
20070221 & 4152 & Gemini-N/GMOS-N  & R400/700~nm/0.75$''$ & 3.5~hrs & 0.76$''$ & 
-- \\
20070222 & 4153 & Gemini-N/GMOS-N  & g' & 600~s & 0.66$''$ & 26.21$^m\pm$0.2$^m$ \\
20070222 & 4153 & Gemini-N/GMOS-N  & r' & 600~s & 0.53$''$ & 23.26$^m\pm$0.15$^m$\\
20070222 & 4153 & Gemini-N/GMOS-N  & i' & 600~s & 0.59$''$ & 21.14$^m\pm$0.1$^m$ \\
20070222 & 4153 & Gemini-N/GMOS-N  & z' & 600~s & 0.63$''$ & 19.05$^m\pm$0.1$^m$ \\
20070222 & 4153 & IRTF/SpeX & SXD,LXD/0.8$''$ & 1600~s,1200~s & 0.90$''$ & 
-- \\
20070318 & 4177 & Gemini-N/Michelle & N'(11.2~$\mu$m) & 100~s & 0.48$''$ & 
0.23$\pm$0.02~Jy \\
20070318 & 4177 & Gemini-N/Michelle & Qa(18.5~$\mu$m) & 100~s & 0.66$''$ & 
0.44$\pm$0.05~Jy \\
20070405 &  & Gemini-N/Michelle & lowN/0.4$''$ & 800~s & 0.80$''$ & 
-- \\
\enddata

\tablenotetext{a}{Modified Julian Date. 2450000+}
\end{deluxetable}
\end{center}
\clearpage


\begin{deluxetable}{ccccc}
\tabletypesize{\scriptsize}
\tablecaption{Optical and Near-IR Equivalent Widths \& Line Fluxes\label{lineflux}}
\tablewidth{0pc}
\tablehead{
\colhead{Line/Band} & 
\colhead{Wavelength} & 
\colhead{W$_\lambda$\tablenotemark{a}} &
\colhead{Observed Line Flux\tablenotemark{b}} &
\colhead{Dereddened Line Flux\tablenotemark{c}} \\
\colhead{} &
\colhead{(~$\mu$m)} & 
\colhead{(\AA)} &
\colhead{(W~m$^{-2}$)} &
\colhead{(W~m$^{-2}$)}}

%

\startdata
O~I          & 0.5579 & -104.4 & 4.7~10$^{-19}$ & 1.3~10$^{-11}$ \\
H$\alpha$    & 0.6563 & -128.4 & 1.0~10$^{-18}$ & 5.4~10$^{-13}$ \\
$[$S~II$]$   & 0.6716 & -16.3  & 1.8~10$^{-19}$ & 6.0~10$^{-14}$ \\
$[$S~II$]$   & 0.6731 & -22.0  & 2.3~10$^{-19}$ & 7.3~10$^{-14}$ \\
Fe~I         & 0.7912 & -7.2   & 4.0~10$^{-19}$ & 7.1~10$^{-15}$ \\
O~I          & 0.8449 & -5.0   & 3.8~10$^{-19}$ & 2.5~10$^{-15}$ \\  
Fe~I         & 0.8467 & -1.9   & 1.7~10$^{-19}$ & 1.1~10$^{-15}$ \\
Ca~II        & 0.8498 & -33.3  & 2.9~10$^{-18}$ & 1.8~10$^{-14}$ \\
Ca~II        & 0.8542 & -37.5  & 3.4~10$^{-18}$ & 1.9~10$^{-14}$ \\
Fe~II        & 0.8600 & -2.7   & 2.6~10$^{-19}$ & 1.4~10$^{-15}$ \\
$[$Fe~II$]$  & 0.8617 & -3.0   & 2.7~10$^{-19}$ & 1.4~10$^{-15}$ \\
Ca~II        & 0.8662 & -35.1  & 3.4~10$^{-18}$ & 1.6~10$^{-14}$ \\
Fe~I         & 0.8689 & -1.6   & 1.6~10$^{-19}$ & 7.3~10$^{-16}$ \\
Pa~12        & 0.8748 & -3.1   & 3.2~10$^{-19}$ & 1.3~10$^{-15}$ \\
Pa~11        & 0.8860 & -4.2   & 4.6~10$^{-19}$ & 1.6~10$^{-15}$ \\
Pa~10        & 0.9012 & -4.7   & 4.9~10$^{-19}$ & 1.4~10$^{-15}$ \\
He~I         & 1.0830 & -10.0  & 2.0~10$^{-18}$ & 7.6~10$^{-16}$ \\
Pa$\gamma$   & 1.0941 & -23.0  & 3.5~10$^{-18}$ & 1.2~10$^{-15}$ \\
$[$Fe~II$]$  & 1.2570 & -2.0   & 7.0~10$^{-19}$ & 7.4~10$^{-17}$ \\
Pa$\beta$    & 1.2822 & -25.0  & 8.6~10$^{-18}$ & 7.9~10$^{-16}$ \\
$[$Fe~II$]$  & 1.5335 & -2.0   & 2.3~10$^{-18}$ & 6.8~10$^{-17}$ \\
$[$Fe~II$]$  & 1.6445 & -2.0   & 2.8~10$^{-18}$\tablenotemark{d} & 5.8~10$^{-17}$ \\
Br~10-4      & 1.7367 & -3.0   & 6.2~10$^{-18}$ & 1.0~10$^{-16}$ \\
Br$\delta$   & 1.9451 & -3.0   & 1.0~10$^{-17}$ & 1.0~10$^{-16}$ \\
v=1-0~S(1)   & 2.1218 & -1.0   & 5.2~10$^{-18}$ & 3.9~10$^{-17}$ \\
Br$\gamma$   & 2.1661 & -5.0   & 2.3~10$^{-17}$ & 1.6~10$^{-16}$ \\
Na~I         & 2.2060 & 0.58   & absorption & absorption \\
Ca~I         & 2.2650 & 0.49   & absorption & absorption \\
v=2-0 CO     & 2.2935 & 1.90   & absorption & absorption \\
Br$\alpha$   & 4.0522 & -12.0  & 7.1~10$^{-17}$  & 1.5~10$^{-16}$ \\
\enddata

\tablenotetext{a}{W$_\lambda$ is the line equivalent width. Negative W$_\lambda$ are emission, positive W$_\lambda$ are absorption.}
\tablenotetext{b}{Emission line fluxes. Uncertainties are estimated at $\sim\pm$10\%}
\tablenotetext{c}{Emission line fluxes dereddened by A$_V$=19$^m$ using A$_{\lambda}$=A$_V$(0.55/$\lambda$($\mu m$))$^{1.6}$}
\tablenotetext{d}{Possible inverse P~Cygni profile.}
\end{deluxetable}
\clearpage

\clearpage
\clearpage
\clearpage
\clearpage
\clearpage
\clearpage
\clearpage
\clearpage
\clearpage
\clearpage
\clearpage
\clearpage
\clearpage
\clearpage
\clearpage
\clearpage
\clearpage
\clearpage
\clearpage


\begin{thebibliography}{}

\bibitem[Acosta-Pulido et al.(2007)]{2007AJ....133.2020A}   Acosta-Pulido, J.~A., et al.\ 2007, \aj, 133, 2020 (AP07)

\bibitem[{\'A}brah{\'a}m et al.(2004)]{2004A&A...419L..39A}   \'Abrah\'am, P., K{\'o}sp{\'a}l, {\'A}., Csizmadia, S.,   Mo{\'o}r, A., Kun, M., \& Stringfellow, G.\ 2004, \aap, 419, L39

\bibitem[Andrews et al.(2004)]{2004ApJ...610L..45A} Andrews, S.~M., 
Rothberg, B., \& Simon, T.\ 2004, \apjl, 610, L45 

\bibitem[Aspin et al.(2006)]{2006AJ....132.1298A} Aspin, C., Barbieri,   C., Boschi, F., Di Mille, F., Rampazzi, F., Reipurth, B., \&   Tsvetkov, M.\ 2006, \aj, 132, 1298

\bibitem[Bacciotti(2002)]{2002RMxAC..13....8B} Bacciotti, F.\ 2002,   Rev. Mex. Astron. Astrofis Conference Series, v13, p8

\bibitem[Baker \& Menzel(1938)]{1938ApJ....88...52B} Baker, J.~G., \& 
Menzel, D.~H.\ 1938, \apj, 88, 52 

\bibitem[Baraffe et al.(1998)]{1998A&A...337..403B} Baraffe, I.,   Chabrier, G., Allard, F., \& Hauschildt, P.~H.\ 1998, \aap, 337, 403

\bibitem[Beck(2007)]{2007AJ....133.1673B} Beck, T.~L.\ 2007, \aj, 133, 1673 

\bibitem[Beck et al.(2007)]{2007AJ....133.1221B} Beck, T.~L., Riera,   A., Raga, A.~C., \& Reipurth, B.\ 2007, \aj, 133, 1221

\bibitem[Bell \& Lin(1994)]{1994ApJ...427..987B} Bell, K.~R., \& Lin, 
D.~N.~C.\ 1994, \apj, 427, 987 

\bibitem[Brice{\~n}o et al.(2004)]{2004ApJ...606L.123B} Brice{\~n}o,   C., et al.\ 2004, \apjl, 606, L123

\bibitem[Bonnell \& Bastien(1992)]{1992ApJ...401L..31B} Bonnell,   I. \& Bastien, P.\ 1992, \apjl, 401, L31

\bibitem[Buehrke et al.(1988)]{1988A&A...200...99B} B\"uhrke, T.,   Mundt, R., \& Ray, T.~P.\ 1988, \aap, 200, 99

\bibitem[Carpenter et al.(2001)]{2001AJ....121.3160C} Carpenter,   J.~M., Hillenbrand, L.~A., \& Skrutskie, M.~F.\ 2001, \aj, 121, 3160

\bibitem[Cohen \& Jones(1987)]{1987ApJ...321..846C} Cohen, M. \&   Jones, B.~F.\ 1987, \apj, 321, 846

\bibitem[Cushing et al.(2004)]{2004PASP..116..362C} Cushing, M.~C.,   Vacca, W.~D., \& Rayner, J.~T.\ 2004, \pasp, 116, 362

\bibitem[Davies et al.(1997)]{1997SPIE.2871.1099D} Davies, R.~L.,   et al.\ 1997, \procspie, 2871, 1099

\bibitem[Dopita(1978)]{1978ApJS...37..117D} Dopita, M.~A.\ 1978, \apjs, 37, 117 

\bibitem[Draine(2003)]{2003ARA&A..41..241D} Draine, B.~T.\ 2003, \araa,   41, 241

\bibitem[Eisloffel \& Mundt(1997)]{1997AJ....114..280E} Eisl\"offel, J.   \& Mundt, R.\ 1997, \aj, 114, 280

\bibitem[Fedele et al.(2007)]{2007arXiv0707.0672F} Fedele, D., van den 
Ancker, M.~E., Petr-Gotzens, M.~G., Ageorges, N., \& Rafanelli, P.\ 2007, ArXiv e-prints, 707, arXiv:0707.0672 (a)

\bibitem[Fedele et al.(2007)]{2007arXiv0706.3281F} Fedele, D., van den 
Ancker, M.~E., Petr-Gotzens, M.~G., \& Rafanelli, P.\ 2007, ArXiv e-prints, 706, arXiv:0706.3281 (b)

\bibitem[Gerakines et al.(1995)]{1995A&A...296..810G} Gerakines, P.~A., 
Schutte, W.~A., Greenberg, J.~M., \& van Dishoeck, E.~F.\ 1995, \aap, 296, 
810 

\bibitem[Gerakines et al.(1996)]{1996A&A...312..289G} Gerakines, P.~A., 
Schutte, W.~A., \& Ehrenfreund, P.\ 1996, \aap, 312, 289 

\bibitem[Gibb et al.(2006)]{2006ApJ...641..383G} Gibb, E.~L., Rettig,   T.~W., Brittain, S.~D., Wasikowski, D., Simon, T., Vacca, W.~D.,   Cushing, M.~C., \& Kulesa, C.\ 2006, \apj, 641, 383

\bibitem[Goodrich(1987)]{1987PASP...99..116G} Goodrich, R.~W.\   1987, \pasp, 99, 116

\bibitem[Gredel(1996)]{1996A&A...305..582G} Gredel, R.\ 1996, \aap, 305, 
582 

\bibitem[Gredel(1994)]{1994A&A...292..580G} Gredel, R.\ 1994, \aap, 292, 
580 

\bibitem[Green et al.(2006)]{2006ApJ...648.1099G} Green, J.~D.,   Hartmann, L., Calvet, N., Watson, D.~M., Ibrahimov, M., Furlan, E.,   Sargent, B., \& Forrest, W.~J.\ 2006, \apj, 648, 1099

\bibitem[Greene \& Lada(1997)]{1997AJ....114.2157G} Greene, T.~P., \&   Lada, C.~J.\ 1997, \aj, 114, 2157

\bibitem[Gullbring et al.(1998)]{1998ApJ...492..323G} Gullbring, E., 
Hartmann, L., Brice\~no, C., \& Calvet, N.\ 1998, \apj, 492, 323 

\bibitem[Grosso et al.(2005)]{2005A&A...438..159G} Grosso, N.,   Kastner, J.~H., Ozawa, H., Richmond, M., Simon, T., Weintraub,   D.~A., Hamaguchi, K., \& Frank, A.\ 2005, \aap, 438, 159

\bibitem[Hall et al.(2004)]{2004SPIE.5499....1H} Hall, D.~N.~B.,   Luppino, G., Hodapp, K.~W., Garnett, J.~D., Loose, M., \&   Zandian, M.\ 2004, \procspie, 5499, 1

\bibitem[Hamann \& Persson(1990)]{1990ASPC....9..304H} Hamann, F. \& 
Persson, S.~E.\ 1990, Cool Stars, Stellar Systems, and the Sun, v9, p304 

\bibitem[Hamann \& Persson(1992)]{1992ApJS...82..247H} Hamann, F. \& Persson, S.~E.\ 1992, \apjs, 82, 247

\bibitem[Hamann et al.(1994)]{1994ApJ...436..292H} Hamann, F., Simon, M., 
Carr, J.~S., \& Prato, L.\ 1994, \apj, 436, 292 

\bibitem[Hartmann et al.(1996)]{1996ApJ...464..387H} Hartmann, L., Calvet, 
N., \& Boss, A.\ 1996, \apj, 464, 387 

\bibitem[Hartmann et al.(1998)]{1998ApJ...495..385H} Hartmann, L.,   Calvet, N., Gullbring, E., \& D'Alessio, P.\ 1998, \apj, 495, 385

\bibitem[Hartmann \& Kenyon(1985)]{1985ApJ...299..462H} Hartmann,   L. \& Kenyon, S.~J.\ 1985, \apj, 299, 462

\bibitem[Hartmann \& Kenyon(1996)]{1996ARA&A..34..207H} Hartmann,   L. \& Kenyon, S.~J.\ 1996, \araa, 34, 207

\bibitem[Heathcote \& Reipurth(1992)]{1992AJ....104.2193H} Heathcote,   S. \& Reipurth, B.\ 1992, \aj, 104, 2193

\bibitem[Herbig(1977)]{1977ApJ...217..693H} Herbig, G.~H.\ 1977,   \apj, 217, 693

\bibitem[Herbig(1989)]{1989lmsf.conf..233H} Herbig, G.~H.\ 1989, in ESO Workshop on Low Mass Star Formation and Pre-main Sequence Objects, ed. Bo Reipurth, p233

\bibitem[Herbig et al.(2001)]{2001PASP..113.1547H} Herbig, G.~H.,   Aspin, C., Gilmore, A.~C., Imhoff, C.~L., \& Jones, A.~F.\ 2001,   \pasp, 113, 1547

\bibitem[Herbig \& Soderblom(1980)]{1980ApJ...242..628H} Herbig,   G.~H. \& Soderblom, D.~R.\ 1980, \apj, 242, 628

\bibitem[Hummer \& Storey(1987)]{1987MNRAS.224..801H} Hummer, D.~G. \& 
Storey, P.~J.\ 1987, \mnras, 224, 801 

\bibitem[Kastner et al.(2004)]{2004Natur.430..429K} Kastner, J.~H., et   al.\ 2004, \nat, 430, 429

\bibitem[Kastner et al.(2006)]{2006ApJ...648L..43K} Kastner, J.~H., et   al.\ 2006, \apjl, 648, L43

\bibitem[Kenyon et al.(1993)]{1993ApJ...414..773K} Kenyon, S.~J., Whitney, 
B.~A., Gomez, M., \& Hartmann, L.\ 1993, \apj, 414, 773 

\bibitem[Kenyon \& Hartmann(1995)]{1995ApJS..101..117K} Kenyon, S.~J.   \& Hartmann, L.\ 1995, \apjs, 101, 117

\bibitem[Kospal et al.(2005)]{2005IBVS.5661....1K} K\'osp\'al, A.,   Abraham, P.~A.-P.~J., Csizmadia, S., Eredics, M., Kun, M., \& Racz,   M.\ 2005, Informational Bulletin on Variable Stars, 5661

\bibitem[Lehmann et al.(1995)]{1995A&A...300L...9L} Lehmann, T.,   Reipurth, B., \& Brandner, W.\ 1995, \aap, 300, L9

\bibitem[Lis et al.(1999)]{1999ApJ...527..856L} Lis, D.~C., Menten,   K.~M., \& Zylka, R.\ 1999, \apj, 527, 856

\bibitem[McGehee et al.(2004)]{2004ApJ...616.1058M} McGehee, P.~M.,   Smith, J.~A., Henden, A.~A., Richmond, M.~W., Knapp, G.~R.,   Finkbeiner, D.~P., Ivezi{\'c}, {\v Z}., \& Brinkmann, J.\ 2004,   \apj, 616, 1058

\bibitem[McNeil (2004)]{2004IAUC.8284....1M} McNeil, J.~W.\ 2004,   \iaucirc, 8284

\bibitem[Meyer et al.(1997)]{1997AJ....114..288M} Meyer, M.~R., Calvet,   N., \& Hillenbrand, L.~A.\ 1997, \aj, 114, 288

\bibitem[Mosoni et al.(2005)]{2005AN....326..565M} Mosoni, L., Ratzka,   T., Abraham, P., Kospal, A., \& Henning, T.\ 2005, Astron.   Nach., 326, 565

\bibitem[Muzerolle et al.(1998)]{1998AJ....116.2965M} Muzerolle, J., 
Hartmann, L., \& Calvet, N.\ 1998, \aj, 116, 2965 

\bibitem[Muzerolle et al.(2001)]{2001ApJ...550..944M} Muzerolle, J., 
Calvet, N., \& Hartmann, L.\ 2001, \apj, 550, 944 

\bibitem[Muzerolle et al.(2003)]{2003ApJ...592..266M} Muzerolle, J.,   Hillenbrand, L., Calvet, N., Brice{\~n}o, C., \& Hartmann, L.\ 2003,   \apj, 592, 266

\bibitem[Muzerolle et al.(2005)]{2005ApJ...620L.107M} Muzerolle, J.,   Megeath, S.~T., Flaherty, K.~M., Gordon, K.~D., Rieke, G.~H., Young,   E.~T., \& Lada, C.~J.\ 2005, \apjl, 620, L107

\bibitem[Natta et al.(2004)]{2004A&A...424..603N} Natta, A., Testi, L.,   Muzerolle, J., Randich, S., Comer{\'o}n, F., \& Persi, P.\ 2004,   \aap, 424, 603

\bibitem[Ojha et al.(2004)]{2004IAUC.8306....2O} Ojha, D.~K.,   Kusakabe, N., \& Tamura, M.\ 2004, \iaucirc, 8306

\bibitem[Ojha et al.(2005)]{2005BASI...33..370O} Ojha, D.~K., et al.\  2005, Bull. Astron. Soc. India, 33, 370

\bibitem[Ojha et al.(2006)]{2006MNRAS.368..825O} Ojha, D.~K., et al.\   2006, \mnras, 368, 825

\bibitem[Pollack et al.(1994)]{1994ApJ...421..615P} Pollack, J.~B., 
Hollenbach, D., Beckwith, S., Simonelli, D.~P., Roush, T., \& Fong, W.\ 1994, \apj, 421, 615

\bibitem[Prato et al.(2003)]{2003ApJ...584..853P} Prato, L., Greene,   T.~P., \& Simon, M.\ 2003, \apj, 584, 853

\bibitem[Quanz et al.(2007)]{2007arXiv0706.3593Q} Quanz, S.~P., Henning, 
T., Bouwman, J., van Boekel, R., Juhasz, A., Linz, H., Pontoppidan, K.~M., 
\& Lahuis, F.\ 2007, ArXiv e-prints, 706, arXiv:0706.3593 

\bibitem[Rayner et al.(2003)]{2003PASP..115..362R} Rayner, J.~T.,   Toomey, D.~W., Onaka, P.~M., Denault, A.~J., Stahlberger, W.~E.,   Vacca, W.~D., Cushing, M.~C., \& Wang, S.\ 2003, \pasp, 115, 362

\bibitem[Reipurth(1989)]{1989Natur.340...42R} Reipurth, B.\ 1989, \nat, 
340, 42 

\bibitem[Reipurth \& Aspin(1997)]{1997AJ....114.2700R} Reipurth,   B. \& Aspin, C.\ 1997, \aj, 114, 2700

\bibitem[Reipurth et al.(2000)]{2000AJ....120.1449R} Reipurth, B., Yu, 
K.~C., Heathcote, S., Bally, J., \& Rodr{\'{\i}}guez, L.~F.\ 2000, \aj, 
120, 1449 

\bibitem[Reipurth \& Aspin(2004)]{2004ApJ...606L.119R} Reipurth, B.,   \& Aspin, C.\ 2004, \apjl, 606, L119

\bibitem[Rettig et al.(2005)]{2005ApJ...626..245R} Rettig, T.~W.,   Brittain, S.~D., Gibb, E.~L., Simon, T., \& Kulesa, C.\ 2005, \apj,   626, 245

\bibitem[Semkov(2004)]{2004IBVS.5578....1S} Semkov, E.~H.\ 2004,   Informational Bulletin on Variable Stars, 5578

\bibitem[Semkov(2006)]{2006IBVS.5683....1S} Semkov, E.~H.\ 2006,   Informational Bulletin on Variable Stars, 5683

\bibitem[Shiba et al.(1993)]{1993ApJS...89..299S} Shiba, H., Sato, S., 
Yamashita, T., Kobayashi, Y., \& Takami, H.\ 1993, \apjs, 89, 299 

\bibitem[Siess et al.(2000)]{2000A&A...358..593S} Siess, L., Dufour,   E., \& Forestini, M.\ 2000, \aap, 358, 593

\bibitem[Smith et al.(1989)]{1989ApJ...344..413S} Smith, R.~G.,   Sellgren, K., \& Tokunaga, A.~T.\ 1989, \apj, 344, 413

\bibitem[Stark et al.(2006)]{2006ApJ...649..900S} Stark, D.~P.,   Whitney, B.~A., Stassun, K., \& Wood, K.\ 2006, \apj, 649, 900

\bibitem[Teixeira \& Emerson(1999)]{1999A&A...351..292T} Teixeira,   T.~C. \& Emerson, J.~P.\ 1999, \aap, 351, 292

\bibitem[Tokunaga \& Vacca(2005)]{2005PASP..117.1459T} Tokunaga,   A.~T. \& Vacca, W.~D.\ 2005, \pasp, 117, 1459

\bibitem[Tsukagoshi et al.(2005)]{2005PASJ...57L..21T} Tsukagoshi, T.,   Kitamura, Y., Kawabe, R., Saito, M., Yokogawa, S., \& Kurono, Y.\   2005, \pasj, 57, L21

\bibitem[Vacca et al.(2003)]{2003PASP..115..389V} Vacca, W.~D.,   Cushing, M.~C., \& Rayner, J.~T.\ 2003, \pasp, 115, 389

\bibitem[Vacca et al.(2004)]{2004ApJ...609L..29V} Vacca, W.~D.,   Cushing, M.~C., \& Simon, T.\ 2004, \apjl, 609, L29

\bibitem[Vig et al.(2006)]{2006A&A...446.1021V} Vig, S., Ghosh, S.~K.,   Kulkarni, V.~K., \& Ojha, D.~K.\ 2006, \aap, 446, 1021

\bibitem[Walter et al.(2004)]{2004AJ....128.1872W} Walter, F.~M.,   Stringfellow, G.~S., Sherry, W.~H., \& Field-Pollatou, A.\ 2004,   \aj, 128, 1872

\bibitem[Weingartner \& Draine(2001)]{2001ApJ...548..296W} Weingartner, 
J.~C. \& Draine, B.~T.\ 2001, \apj, 548, 296 

\bibitem[White \& Basri(2003)]{2003ApJ...582.1109W} White, R.~J. \&   Basri, G.\ 2003, \apj, 582, 1109

\bibitem[Whitney \& Hartmann(1993)]{1993ApJ...402..605W} Whitney, B.~A. \& 
Hartmann, L.\ 1993, \apj, 402, 605 

\end{thebibliography}
\end{document}